\documentclass[11pt]{article}
\usepackage{mathrsfs}
\usepackage[centertags]{amsmath}
\usepackage{amsfonts}
\usepackage{amssymb}
\usepackage{amsthm}
\usepackage{cases}
\usepackage{indentfirst}
\usepackage{epsfig}
\usepackage {graphicx}
\usepackage{color}
\usepackage{CJK}
\usepackage{hyperref}
\usepackage{setspace}
\usepackage{algorithm}
\usepackage{algorithmicx}
\usepackage{algpseudocode}
\usepackage{natbib}
\usepackage{soul}
\usepackage{xr}
\usepackage{comment}

\date{}

\newtheorem{theorem}{Theorem}[section]

\newtheorem{lemma}{Lemma}[section]
\newtheorem{condition}{Condition}[section]

\newtheorem{remark}{Remark}[section]

\numberwithin{equation}{section}
\renewcommand{\theequation}{\thesection.\arabic{equation}}

\def\E{\operatorname*{E}}

\def\Cov{\operatorname*{Cov}}
\def\Cor{\operatorname*{Cor}}
\def\sign{\operatorname*{sign}}

\def\Rank{\operatorname*{Rank}}
\def\diag{\operatorname*{diag}}
\def\swap{\operatorname*{swap}}
\def\FDR{\operatorname*{FDR}}
\def\Pr{\operatorname*{Pr}}
\def\D{\operatorname*{D}}

\allowdisplaybreaks[4]

\newcommand{\bbeta}{\boldsymbol{\beta}}
\def\bd{\mathbf{d}}
\def\hbtheta{\hat{\boldsymbol{\theta}}}
\def\bdelta{\boldsymbol{\delta}}
\def\hbtheta{\hat{\boldsymbol{\theta}}}
\def\tbtheta{\tilde{\boldsymbol{\theta}}}

\def\by{\mathbf{y}}
\def\bX{\mathbf{X}}
\def\beps{\boldsymbol{\epsilon}}
\def\btheta{\boldsymbol{\theta}}
\def\bM{\mathbf{M}}
\def\bI{\mathbf{I}}
\def\tbX{\tilde{\mathbf{X}}}
\def\bSigma{\boldsymbol{\Sigma}}
\def\bs{\mathbf{s}}

\def\tbtheta{\tilde{\boldsymbol{\theta}}}

\def\bZ{\mathbf{Z}}

\def\supp{\operatorname{supp}}

\def\hsigma{\hat{\sigma}}

\def\calS{\mathcal{S}}

\def\hbbeta{\hat{\boldsymbol{\beta}}}

\def\hcalS{\hat{\mathcal{S}}}
\def\hbeta{\hat{\beta}}

\def\ttheta{\tilde{\theta}}
\def\htheta{\hat{\theta}}

\begin{document}
\title{\textbf{A Generalized Knockoff Procedure for FDR Control in Structural Change Detection}}
\author{Jingyuan Liu$^{a}$,  Ao Sun$^{a}$ and Yuan Ke$^{b}$   \\
$^{a}$MOE Key Laboratory of Econometrics, Department of Statistics, \\
                 School of Economics, Wang Yanan Institute for Studies in Economics \\
                 and Fujian Key Lab of Statistics, Xiamen University, P.R China\\
$^{b}$Department of Statistics,  University of Georgia, Athens, GA 30602, USA}

\maketitle
\begin{quote}
	{\bf Abstract.} Controlling false discovery rate (FDR) is crucial for variable selection, multiple testing, among other signal detection problems. In literature, there is certainly no shortage of FDR control strategies when selecting individual features, but the relevant works for structural change detection, such as profile analysis for piecewise constant coefficients and integration analysis with multiple data sources, are limited.
	In this paper, we propose a generalized knockoff procedure (GKnockoff) for FDR control under such problem settings. We prove that the GKnockoff possesses pairwise exchangeability, and is capable of controlling the exact FDR under finite sample sizes. We further explore GKnockoff under high dimensionality, by first introducing a new screening method to filter the high-dimensional potential structural changes. We adopt a data splitting technique to first reduce the dimensionality via screening and then conduct GKnockoff on the refined selection set. Furthermore, the powers of proposed methods are systematically studied. Numerical comparisons with other methods show the superior performance of GKnockoff, in terms of both FDR control and power. We also implement the proposed methods to analyze a macroeconomic dataset for detecting changes of driven effects of economic development on the secondary industry.

\end{quote}

\section{Introduction}
The era of information explosion has driven researchers from squeezing limited data to extracting useful messages from massive amounts of data. Plentiful works have been developed for detecting important features, ranging from regularized regression \citep{Tibshirani, FanLi2001,Zhang2010nearly, Fan2020factor} to screening-related approaches \citep{Fan2008,Li2012,Liu2014feature,Mai2015, Ma2017variable, Liu2020}. See \cite{fan2014sure}, \cite{Liu2015} and \cite{fan2020statistical} for summaries of important works among those lines. Meanwhile, apart from identifying individual features, structural change detection is also of great scientific interest, especially in the realm of finance, genomics, health care, social science, and so forth. For instance, identifying the impact of economic structural changes is a crucial task in the macroeconomic study since the structural changes might alter economic assumptions for determining courses of action \citep{ramey2016macroeconomic}. The structural changes non-exhaustively include effect changes in piecewise constant coefficient models, and heterogenous coefficients upon integrating multiple data sources. \cite{ke2015homogeneity} proposed a CARDS method to first order the coefficients and then fuse the adjacent coefficients. \cite{chen2015graph} studied a graph-based change point detection method. \cite{wang2016fused} applied the CARDS idea to combine multiple studies with repeated measurements to reduce constraints on coefficients and to gain computational efficiency. \cite{Tang2010} utilized fused Lasso \citep{Tibshirani2005Sparsity} to identify the heterogeneous coefficients by merging inter-study homogeneous parameter clusters. More recent developments include \cite{avanesov2018change}, \cite{wang2020optimal}, \cite{jiang2020time}, \cite{dette2020estimating}, \cite{ke2021homogeneity}, among many others.

Admittedly, researchers have devoted much attention to detecting structural changes.  Another crucial question to consider, however, is how ``scientific" the discoveries are - if the extracted information consists of too many falsely discovered signals that are merely selected to fit the current random sample,
we would be stuck in the ``reproducible crisis" that deeply undermines the reliability of statistical findings. The false discoveries are attributed in part to the classical error accumulation issue raised by multiple testing, and the spurious correlations \citep{Fan2008, Fan2012variance}. The spurious correlation between a certain feature and the response refers to the respective association that is exhibited merely by the current specific sample, rather than the nature of the population relationship. For instance, a spurious feature might seem to be predictive to the response because it is correlated with some true predictors. Unfortunately, such spuriousness can not be revealed by standard signal detection techniques. For feature selection problem, \cite{Su2017} demonstrated that the true and null features often intersperse on the Lasso solution path when all features are independently generated from an identical Gaussian distribution, and this phenomenon occurs under whatsoever effect sizes. Then directly selecting nonzero coefficients along the solution path fails to control false discoveries.

Accordingly, there is an eager appeal of controlling the degree of false discoveries, typically measured by the false discovery rate (FDR), to enhance the reliability of structural change detection. The concept of FDR was first advocated by \cite{Benjamini1995}, calculated as the expected proportion of false discoveries among all discoveries. To be specific, let $\mathcal{S}$ be the index set of the true signals, and $\hat{\mathcal{S}}$ be that of the discovered signals based on the sampled data. The FDR is defined as
\begin{equation}
	\FDR =  \E \left[ \frac{|\hat{\mathcal{S}} \setminus \mathcal{S}|}{|\hat{\mathcal{S}}|}\right].
	\label{equ:fdr}
\end{equation}

For independent multiple testing problems, \cite{Benjamini1995} developed a sequential Bonferroni-type method, called the B-H procedure, to control FDR based on the ordered individual p-values. \cite{Benjamini2001} showed that the B-H procedure also works under the assumption of ``positive regression dependence on a subset". They further advocated a B-Y method by adding a divisor to the threshold of B-H, and proved that B-Y can control FDR under arbitrary dependence structures. Refer to \cite{benjamini2010discovering} for a comprehensive overview of the B-H-type methods. A more recent milestone of FDR control is the proposal of knockoff filter \citep{Rina2015}. It constructs ``knockoffs" for the original features that mimic the correlation structure among features yet are known to be independent of the response. Then the knockoffs might serve as references for estimating and hence controlling FDR by regressing the response on both original and knockoff features. Under mild conditions, the knockoff filter achieves exact FDR control in finite sample settings. \cite{Dai2016} extended the knockoff filter to grouped feature selection. \cite{candes2018} developed a model-X knockoff procedure that can be applied to the high-dimensional regime, provided the prior knowledge about the joint distributions of the original features. \cite{fan2019rank} constructed the model-X knockoffs when the covariates follow a Gaussian graphical model. \cite{lu2018} integrated the model-X knockoff framework with the deep neural networks (DNN) architecture to enhance the interpretability and reproducibility of DNN. If the joint distribution of features is unknown, \cite{fan2020ipad} and \cite{romano2020deep} constructed the knockoff variables by imposing a latent factor model and a deep generative model, respectively.

As the other side of the coin, the power of knockoff filter has also been systematically studied. \cite{weinstein2017power} proved that the knockoff filter from an independent and identically distributed Gaussian design asymptotically achieves optimal power. \cite{fan2019rank} proved that the model-X knockoff achieves optimal power asymptotically for the linear model with independent sub-gaussian noises. \cite{ke2020power} systematically analyzed the power of knockoffs under the rare and weak signal regimes and derived the FDR-TPR(True Positive Rate) trade-off diagram. In sum, the knockoff filter has been shown to possess comparable selection power with many other FDR control methods when the samples are independent.

Notwithstanding the merit of knockoffs, few related literature, to our best knowledge, is amenable to the  structural change detection. Thus in this paper, we propose a unified approach, called generalized knockoff (GKnockoff), that rigorously controls FDR for structural change problems under finite sample sizes. Consider the classical linear regression model
\begin{equation}
	\mathbf{y} = \mathbf{X} \boldsymbol{\beta} + \boldsymbol{\epsilon},
	\label{equ:main}
\end{equation}
where $\mathbf{y} \in \mathbb{R}^n$ is a sample vector of response, $\mathbf{X} \in \mathbb{R}^{n \times p}$ is a design matrix, $\boldsymbol{\beta}=(\beta_1,\ldots,\beta_p)^\top\in \mathbb{R}^p$ is an unknown vector of coefficients, and $\boldsymbol{\epsilon}$ is independent with $\mathbf{X}$. Further, we assume the elements in $\boldsymbol{\epsilon}$  are i.i.d. normal random errors with mean 0 and variance $\sigma^2$.

While variable selection identifies nonzero $\beta_j$'s,  structural change detection typically concerns about the linear combinations, $\mathbf{d}_j^\top \boldsymbol{\beta}, \ j=1,\ldots,m$, for some properly defined $\mathbf{d}_j \in \mathbb{R}^{p}$, where $m$ is the number of linear combinations under consideration. Take piecewise constant coefficients profile for instance, it is of interest to detect $\{j=1,\ldots,p-1: \beta_{j+1} - \beta_{j} \ne 0\}$; namely, $m=p-1$, and $\mathbf{d}_j=(0,\ldots,0,-1,1,0,\ldots,0)^\top$, whose $(j+1)$th element is 1 and $j$th element is $-1$. In addition, variable selection could as well belong to this structural change category by defining $\mathbf{d}_j=(0,\ldots,0,1,0,\ldots,0)^\top$ with 1 appearing in the $j$th position. More examples of structural change are discussed in Section 2. Therefore, the objective is indeed to identify the active set $\mathcal{S} = \{j=1,\ldots,m:  \mathbf{d}_j^\top \boldsymbol{\beta} \ne 0\}$ while controlling FDR.

The main obstacle of constructing knockoffs for such problems is twofold. Firstly, we
apply a full-row-rank transformation matrix to transform the original samples, and then recover the active set $\mathcal{S}$ by a partial regularization method.
However, the transformed samples are no longer independent and hence violate the assumption for the theories in \cite{Rina2015}.
To tackle this challenge, we propose a generalized knockoff procedure, named GKnockoff, for the transformed data.
We show that the GKnockoff variables enjoy the exchangeability without the independence assumption, and prove that GKnockoff can rigorously control FDR. Secondly, The framework of \cite{Rina2015} does not apply to the high-dimensional regime, as the construction of knockoff features requires the sample size to be at least twice the number of features. As a primitive philosophy of quickly reducing dimensionality, screening has been extensively studied over the past decade since the pioneering work of \cite{Fan2008}. However, compared with ubiquitous techniques for screening individual features based on various models, our understanding of how to screen structural changes seems limited, so to speak.
The challenge is in part attributed to that screening individual features is a matter of teasing apart signals from noise, thus it only requires effects of active features to be non-vanishingly estimated, typically large. Meanwhile, screening structural changes,   e.g. coefficient changes, which aims to discover the ``difference in coefficients", calls for the ability to accurately quantify the amplitude of individual coefficients.  Accordingly, we develop a new screening procedure for structural change detection and study its theoretical and empirical performances. Also, we adopt a data splitting technique  \citep[e.g.][]{wasserman2009high, barber2019knockoff, Liu2020} to alternatively filtering structural changes and constructing GKnockoffs on two halves of data. Furthermore, the aforementioned non-independent transformed data bring about significant challenges to the power analysis of GKnockoff. Therefore, the power of proposed GKnockoff method is carefully studied under such dependence structure. We also develop an efficient and user-friendly R package \textit{`GKnockoff'} \footnote{\href{https://github.com/suntiansheng/Gknockoff}{https://github.com/suntiansheng/Gknockoff}} to implement the GKnockoff procedure.

\subsection{Notations}
We introduce the following notations used throughout this paper. Denote $\mathbb{R}$  the set of real numbers. For a set $\mathcal{A}$, $|\mathcal{A}|$ denotes its cardinality.  Given a vector $\mathbf{x} = [x_1, \ \ldots, \ x_d]^{\top} \in \mathbb{R}^d$, we write the vector $l_q$-norm as $\|\mathbf{x}\|_q := \big(\sum_{j=1}^d |x_j|^q\big)^{1/q}$ for $1 \leq q < \infty$ and the vector $l_{\infty}$-norm as $\|\mathbf{x}\|_{\infty} := \max_{1 \leq j \leq d}|x_j|$. Denote $\diag\{\mathbf{x}\}$ a diagonal matrix whose diagonal elements belong to $\mathbf{x}$. For a matrix $\mathbf{A} = \big[ \mathbf{A}_{(k,l)}\big]_{1 \leq k \leq d_1; 1 \leq l \leq d_2} \in \mathbb{R}^{d_1 \times d_2}$, the $\infty$-norm of $\mathbf{A}$ is denoted as {$||| \mathbf{A} |||_{\infty} := \max_{k}\sum_{l=1}^{d_2}|\mathbf{A}_{(k,l)}|$}. The $j$th column of $\mathbf{A}$ is denoted $A_j$, and $\mathbf{A}_{\mathcal{G}}$ refers to the columns of $\mathbf{A}$ with indexes in the set $\mathcal{G}$.
{If $\mathbf{A}$ is a symmetric matrix, $\Lambda_{\max}(\mathbf{A})$ and $\Lambda_{\min}(\mathbf{A})$} are respectively its largest and smallest eigenvalues. {We write $\mathbf{A} \succ 0$ if $\mathbf{A}$ is a positive definite matrix.}
$\mathbf{0}_d$, $\mathbf{0}_{d\times d}$ and $\mathbf{I}_d$ denote the $d$-dimensional vector of zeros, the $d \times d$-dimensional matrix of zeros,  and the $d$-dimensional identity matrix, respectively.
For $a, b \in \mathbb{R}$, we denote $\sign(a)$ the sign function of $a$, and $a\vee b$ the maximum between $a$ and $b$. When necessary, we consider $0/0 = 0$.
For two matrices $\mathbf{A}$ and $\mathbf{B}$ of the same dimension, the operator $[\mathbf{A}, \mathbf{B}]_{\swap(\mathcal{G}) }$ refers to swapping the $jth$ column in $\mathbf{A}$ and the $jth$ column in $\mathbf{B}$ for all $j \in \mathcal{G}$. For two sequance $\{a_n\}$ and $\{b_n\}$, $a_n \gg b_n$ means $a_n/b_n \to \infty$ as $n \to \infty$.

\subsection{Organization of the paper}

The rest of the paper is organized as follows. In Section \ref{sec:frame}, we propose a generalized knockoff (GKnockoff) framework to detect structural changes and control FDR. We define a GKnockoff matrix and discuss the intuition and the implementation of a GKnockoff filter. We also provide the theoretical guarantees of GKnockoff for controlling FDR under finite samples and analyze the power of proposed methods. Section \ref{sec:high} is devoted to the high-dimensional structural change detection and FDR control problem, where we introduce a screening technique named FuSIS and a high-dimensional GKnockoff filter. The superior performance of GKnockoff is empirically verified through several simulation studies in Section \ref{sec:simulation}. In Section \ref{sec:realdata}, we apply the proposed Gknockoff filter to detect structural changes of the secondary industry among different provinces. Section 6 concludes the paper. The proofs of theoretical results, along with some remarks, are presented in the online supplementary material.

\section{A generalized knockoff framework}
\label{sec:frame}
\subsection{Problem setup}
\label{prob state}
As we discussed in the introduction, many structural change detection problems can be formulated by the linear model (\ref{equ:main}) and the $m$ hypotheses as follows.
\begin{align*}
    H_{0j}: \ \mathbf{d}_j^\top\bbeta=0 \quad \text{v.s.} \quad
    H_{1j}: \ \mathbf{d}_j^\top\bbeta \neq 0, \quad j=1,\ldots,m,
\end{align*}
where $\mathbf{d}_j \in \mathbb{R}^{p}$ is a problem-driven transformation vector. Further, we denote $\mathcal{S} = \{j=1,\ldots,m:  \mathbf{d}_j^\top \boldsymbol{\beta} \ne 0\}$ the active set of the structural change detection problem, and $\mathcal{S}^c=\{1,\ldots, m\}\setminus\mathcal{S}$ the inactive set. Let $\mathbf{D} = [\mathbf{d}_1,...,\mathbf{d}_m]^\top \in \mathbb{R}^{m \times p}$. We then estimate the regression coefficients and recover the active set $\mathcal{S}$ simultaneously  by solving a generalized Lasso problem \citep{Ryan2011}
\begin{equation}
	\min_{\mathbf{b} \in \mathbb{R}^{p}} \frac{1}{2n}\|\mathbf{y}-\mathbf{X} \mathbf{b}\|_{2}^{2}+\lambda\|\mathbf{D} \mathbf{b}\|_1,
	\label{equ:gen_obj}
\end{equation}
where $\lambda\geq 0$ is a regularization parameter.  The model setup \eqref{equ:gen_obj} is applicable to a wide range of structural change detection problems.
Next, we discuss the design of the transformation matrix $\mathbf{D}$ in two popular scenarios.

\noindent{\bf Scenario 1: Piecewise constant coefficients profile}

Piecewise constant coefficients profile \citep[or homogeneity pursuit, see][]{ke2015homogeneity} assumes the $p$ coefficients $\beta_1,\ldots,\beta_p$ in model (\ref{equ:main}) can be segmented into $J+1$ groups, such that $\beta_{1}=\ldots=\beta_{\tau_{1}} \neq \beta_{\tau_{1}+1} =\ldots=\beta_{\tau_{2}} \neq \beta_{\tau_{2}+1} =\ldots=\beta_{\tau_{J}} \neq \beta_{\tau_{J}+1} =\ldots=\beta_{p}$. Let $\mathcal{S}=\{\tau_1,\ldots,\tau_J\}$ be the set of all change locations. To recover $\mathcal{S}$ by \eqref{equ:gen_obj}, we can choose set $m=p-1$ and design $\mathbf{D}$ in the format as \eqref{equ:difference_matrix}.
\begin{equation}
	\mathbf{D}=\left[\begin{array}{cccccc}
		-1 & 1 & 0 & \cdots & 0 & 0 \\
		0 & -1 & 1 & \cdots & 0 & 0 \\
		\vdots & \vdots & \vdots & \vdots 
		& \vdots & \vdots \\
		0 & 0 & 0 & \cdots & -1 & 1
	\end{array}\right]_{m \times n}.
	\label{equ:difference_matrix}
\end{equation}

\noindent{\bf Scenario 2: Integration analysis from multiple data sources}

Due to the rapid development of data collecting techniques, it has been attractive yet challenging to integrate high-throughput data from multiple sources into a unified regression framework.
Suppose there are $K$ available data sources. For the $k$th data source, $k=1, \ \ldots, \ K$, we observe a random sample $(\mathbf{y}^{(k)}, \mathbf{X}^{(k)})$ and fit a linear regression model $\mathbf{y}^{(k)} = \mathbf{X}^{(k)} \boldsymbol{\beta}^{ (k)} + \boldsymbol{\epsilon}^{(k)}$, where $\mathbf{X}^{(k)} \in \mathbb{R}^{n_k \times p}$ and $\boldsymbol{\epsilon}^{(k)}\sim N(0, \sigma^2\mathbf{I}_{n_k})$. It is natural to test weather there is a homogeneous structure embedded among the coefficient vectors $\{\boldsymbol{\beta}^{ (1)}, \ \ldots, \ \boldsymbol{\beta}^{ (K)}\}$. To that end, we can formulate \eqref{equ:gen_obj} as
{ 
\begin{equation}
	\min_{\mathbf{b} \in \mathbb{R}^{Kp}} \frac{1}{2n}	\left\| \begin{bmatrix}
		\mathbf{y}^{(1)} \\
		\mathbf{y}^{(2)} \\
		\vdots \\
		\mathbf{y}^{(K)}
	\end{bmatrix} -
	\begin{bmatrix}
		&\mathbf{X}^{(1)} &\ldots & \mathbf{0} \\
		&\vdots & \ddots & \vdots \\
		& \mathbf{0} & \ldots &\mathbf{X}^{(K)}
	\end{bmatrix}
	\begin{bmatrix}
		\mathbf{b}^{(1)} \\
	\mathbf{b}^{(2)} \\
		\vdots \\
	\mathbf{b}^{(K)}
	\end{bmatrix} \right\|_2^2 + \lambda\left\|
	\mathbf{D}
\begin{bmatrix}
	\mathbf{b}^{(1)} \\
	\mathbf{b}^{(2)} \\
	\vdots \\
	\mathbf{b}^{(K)}
\end{bmatrix}
 \right\|_1,
	\label{equ:integration1}
\end{equation}
}
where the transformation matrix $\mathbf{D}$ can be designed as
\begin{equation}
	\mathbf{D} = \begin{bmatrix}
		&\mathbf{I}_p & -\mathbf{I}_p & \mathbf{0}_{p\times p} & \ldots & \mathbf{0}_{p\times p} & \mathbf{0}_{p\times p} \\
		& \mathbf{0}_{p\times p} & \mathbf{I}_p & - \mathbf{I}_p & \ldots  & \mathbf{0}_{p\times p} & \mathbf{0}_{p\times p}\\
		& \vdots & \vdots & \vdots & \vdots  & \vdots  & \vdots\\
		&\mathbf{0}_{p\times p} & \mathbf{0}_{p\times p}&\mathbf{0}_{p\times p} &\ldots &\mathbf{I}_p  & -\mathbf{I}_p
	\end{bmatrix}_{(K-1)p \times Kp}.
	\label{equ:generalized_difference}
\end{equation}

Beyond the above examples, the regularization regression  \eqref{equ:gen_obj} and the transformation matrix $\mathbf{D}$ can be tailored for a wide range of structural change detection problems. Further, our problem setup has the {potential} to be applied to other high-dimensional testing problems where the hypothesis can be characterized by a linear combination of regression coefficients. See \cite{runze2021linear} as a recent study in this direction, among others.

\subsection{Structural change detection}

In this subsection, we study the detection of structural changes by solving the generalized Lasso problem (\ref{equ:gen_obj}). Suppose $\mathbf{D}$ is of full row rank. We first define $\tilde{\mathbf{D}} =  [\mathbf{D}^\top,\mathbf{E}^\top ]^\top \in \mathbb{R}^{p \times p}$ such that $\tilde{\mathbf{D}}$ has full rank, with $\mathbf{E}$ being any matrix in the complementary space of the linear space spanned by $\mathbf{D}$ . We also define $\boldsymbol\theta=(\boldsymbol\theta_1^\top,\boldsymbol\theta_2^\top)^\top=\tilde{\mathbf{D}}\boldsymbol\beta$, where $\boldsymbol\theta_1=(\theta_{11},\ldots,\theta_{1m})^\top=\mathbf{D}\boldsymbol\beta$ and $\boldsymbol\theta_2=\mathbf{E}\boldsymbol\beta$.
Then, the non-zero elements in $\boldsymbol\theta_1$ reflect the structural changes of interest and the active set can be represented as $\mathcal{S}=\{j=1,\ldots, m:  \theta_{1j}\neq 0\}$. Further, denote the inverse matrix of $\tilde{\mathbf{D}}$ as $\tilde{\mathbf{D}}^{-1} = [\mathbf{Z}_{p \times m}, \mathbf{F}_{p \times (p-m)}]$, where $\mathbf{Z}_{p \times m}$ and $\mathbf{F}_{p \times (p-m)}$ stand for the first $m$ and the rest $(p-m)$ columns of $\tilde{\mathbf{D}}^{-1}$ respectively. Therefore, we have $\boldsymbol\beta=\tilde{\mathbf{D}}^{-1}\boldsymbol\theta=\mathbf{Z}\boldsymbol\theta_1+\mathbf{F}\boldsymbol\theta_2$.  With the above preparations, we can reformulate \eqref{equ:gen_obj} as a partial regularization problem
\begin{equation}
	\min_{\boldsymbol\theta=[\boldsymbol\theta_1^\top,\boldsymbol\theta_2^\top]^\top \in \mathbb{R}^{p}} \frac{1}{2n}\|\mathbf{y}- \mathbf{X} \mathbf{Z} \boldsymbol\theta_1 - \mathbf{X} \mathbf{F} \boldsymbol\theta_2 \|_{2}^{2}+\lambda  \| \boldsymbol\theta_1 \|_1,
	\label{equ:fuse2}
\end{equation}
where $\lambda\geq 0$ is a regularization parameter. Notice that, when $\mathbf{D}$ is designed as in \eqref{equ:difference_matrix}, the formualiotn in \eqref{equ:fuse2} becomes a fused Lasso problem \citep{Tibshirani2005Sparsity}.

We then adopt the partial residual technique \citep[e.g.][]{Hsiao2021, ZouLi2008} and transfer \eqref{equ:fuse2} to a Lasso-type problem
\begin{equation}
	\min_{\boldsymbol\theta_1 \in \mathbb{R}^{m}} \frac{1}{2n}\|{\mathbf{y}}^*- {\mathbf{X}}^* \boldsymbol\theta_1 \|_{2}^{2}+\lambda  \| \boldsymbol\theta_1 \|_1,
\label{equ:fuse4}
\end{equation}
where ${\mathbf{y}}^* = \mathbf{M} \mathbf{y}$, ${\mathbf{X}}^* =  \mathbf{M} \mathbf{X} \mathbf{Z}$, and $\mathbf{M}: = \mathbf{I}_n - (\mathbf{X}\mathbf{F}) [(\mathbf{X}\mathbf{F})^\top (\mathbf{X}\mathbf{F})]^{-1} (\mathbf{X}\mathbf{F})^\top$ is a projection matrix. The transformed design matrix $\mathbf{X}^*$ consists of $m$ columns, each of which is associated with a potential structural change.
A key observation from (\ref{equ:fuse4}) is that the elements in ${\mathbf{y}}^*$ are no longer independent since
\begin{equation}
	{\mathbf{y}}^* = \mathbf{M} \mathbf{y} \sim  {N}({\mathbf{X}}^* \boldsymbol{\theta}_1, \sigma^2\mathbf{M}).
	\label{equ:y_bar_dis}
\end{equation}
Such dependence violates the independence assumption imposed for the theoretical analysis of knockoff filter \citep{Rina2015}, and hence calls new methodological and theoretical investigations. In the rest of this subsection, we derive the selection consistency and asymptotic power for the solution of  \eqref{equ:fuse4}. In the next subsection, we introduce a generalized knockoff filter to control FDR under the dependence structure.

We establish selection consistency using the Primal-Dual Witness technique \citep{Wainwright2009}. Without loss of generality, we assume that ${X}^*_j$ is normalized such that {$||X_j^*||_2^2/n  = 1$,  $j = 1,\ldots,m$}, to simplify the presentation in the theoretical analysis. In addition, we impose the following two conditions to pave the way for the selection consistency results presented in Theorem \ref{the:power} below.

\begin{condition}\label{cond1}
$	|||{\mathbf{X}}_{ \mathcal{S}^c}^{*\top}{\mathbf{X}}^*_{\mathcal{S}} ({\mathbf{X}}_{\mathcal{S}}^{*\top} {\mathbf{X}}^*_{\mathcal{S}})^{-1}|||_\infty \le 1-\kappa$ for some constant $\kappa \in (0, 1]$.\end{condition}

\begin{condition}\label{cond2}
$   \Lambda_{\min} ({\mathbf{X}}_{\mathcal{S}}^{*\top} {\mathbf{X}}^*_{\mathcal{S}}/n) \ge C_{\min}$ for some constant $C_{\min}>0$,
\end{condition}

Condition \ref{cond1}, which requires the active and inactive variables not to be too correlated, is a common technique condition considered in previous work of Lasso, see, for example, \cite{zhao2006model} and \cite{Wainwright2009}. Condition \ref{cond2} states the minimum eigenvalue of Gram matrix of the true set is bounded away from zero and thus the Gram matrix is invertible. We remark that although the conditions are imposed on the transformed design matrix ${\mathbf{X}}^*$ instead of original matrix $\mathbf{X}$, ${\mathbf{X}}^*$  is observable since ${\mathbf{X}}^* =  \mathbf{M} \mathbf{X} \mathbf{Z}$, where $\mathbf{M}$ and $\mathbf{Z}$ are explicitly obtained upon determination of matrix $\mathbf{D}$.

\begin{theorem}
Under Conditions \ref{cond1} and \ref{cond2}, suppose $\lambda > \frac{2}{\kappa} \sqrt{\frac{2\sigma^2 \log p}{n}}$ in (\ref{equ:fuse4}), then for some $c_1 >0$, the following statements hold with probability greater than $1 - 4\exp(-c_1n \lambda^2)$.
\begin{itemize}
\item[(a)] The generalized Lasso has a unique solution $\hat{\boldsymbol{\theta}}_1$ with $\mathcal{S}\subset\hat{\mathcal{S}}$, where $\hat{\mathcal{S}}=\{j=1,\ldots,m: \ \hat{\theta}_{1j}\neq0\}$. And the estimate $\hat{\boldsymbol{\theta}}_{1{\mathcal{S}}}$ of the truly non-vanishing coefficient $\boldsymbol{\theta}_{1\mathcal{S}}$ satisfies
	\begin{equation}
		||\hat{\boldsymbol{\theta}}_{1{\mathcal{S}}} - \boldsymbol{\theta}_{1\mathcal{S}}||_\infty \le  g(\lambda),	\end{equation}
where $g(\lambda)=\lambda \left[ |||(\mathbf{X}_{\mathcal{S}}^{*\top}\mathbf{X}^*_\mathcal{S}/n )^{-1}|||_\infty + \frac{4 \sigma}{\sqrt{C_{\min}}}\right]$.
\item[(b)] If we further assume $\min_{j \in \mathcal{S}}(|\boldsymbol{\theta}_{1,j}|) \ge g(\lambda) $, then the generalized Lasso estimator has the correct sign, i.e. $\sign(\hat{\boldsymbol{\theta}}_1) = \sign(\boldsymbol{\theta}_1)$.
\end{itemize}
 \label{the:power}
\end{theorem}
\vspace{-0.2in}

Theorem \ref{the:power} (a) guarantees that the generalized Lasso under the dependence structure enjoys the sure screening property \citep{Fan2008}, and hence a full asymptotic power; and the estimation errors are uniformly bounded above. Theorem \ref{the:power} (b) further implies the selection consistency, thus the asymptotic FDR is zero. Nevertheless, the finite sample FDR control, as to be discussed in the next section, is of more interest for practitioners.

\subsection{Generalized knockoff filter and FDR control }

In this subsection, we introduce a generalized knockoff (GKnockoff) filter and an FDR control procedure. Denote  $\boldsymbol\Sigma^*={\mathbf{X}}^{*\top} {\mathbf{X}}^{*}=  \mathbf{Z}^\top\mathbf{X}^\top\mathbf{M} \mathbf{X} \mathbf{Z}$ the Gram matrix of ${\mathbf{X}}^*$. The $n\times m$ matrix of GKnockoff features $\tilde{\mathbf{X}}$ should satisfy
\begin{equation}
	\tilde{\mathbf{X}}^\top \tilde{\mathbf{X}} = \boldsymbol{\Sigma}^* \succ 0, \mbox{ and }  \tilde{\mathbf{X}}^\top {\mathbf{X}}^* = \boldsymbol{\Sigma}^* - \diag\{\mathbf{s}\} .
	\label{eq:X_tilde_matrix}
\end{equation}
The matrix $\tilde{\mathbf{X}}$ can be considered as a second-order knockoff copy of  ${\mathbf{X}}^*$ for the following reasons. First, given ${\mathbf{X}}^*$, $\tilde{\mathbf{X}}$ is independent of ${\mathbf{y}}^*$ since we do not use the information of ${\mathbf{y}}^*$ in \eqref{eq:X_tilde_matrix}. Second, the Gram matrix remains after column-wise swapping, i.e., $[{\mathbf{X}}^{*}, \tilde{ \mathbf{X}}]_{\swap{(\mathcal{G})}}^\top [{\mathbf{X}}^{*}, \tilde{ \mathbf{X}}]_{\swap{(\mathcal{G})}} = [{\mathbf{X}}^{*}, \tilde{ \mathbf{X}}]^\top [{\mathbf{X}}^{*}, \tilde{ \mathbf{X}}]$ for any $\mathcal{G} \subset \{1,2,\ldots,m\}$. When  $n \ge 2m$, one can compute $\tilde{\mathbf{X}}$  by
\begin{equation}
	\tilde{\mathbf{X}}={\mathbf{X}}^*\left( \mathbf{I}_{m}- \boldsymbol{\Sigma}^{*-1} \diag\{\mathbf{s}\}\right)+\tilde{\mathbf{U}} \mathbf{C}
	\label{eq:GK_compute}
\end{equation}
for some $\mathbf{s}=(s_1,\ldots,s_m)^\top \in \mathbb{R}_{+}^{m}$ satisfying $2 \boldsymbol{\Sigma}^* - \diag(\mathbf{s}) \succeq 0$. Moreover, $\tilde{\mathbf{U}}$ is in the null space of ${\mathbf{X}}^*$, i.e. ${\mathbf{X}}^{*\top} \tilde{\mathbf{U}} = 0$ and $\mathbf{C}$ is the Cholesky decomposition of  $2\diag\{\mathbf{s}\} - \diag\{\mathbf{s}\} \boldsymbol{\Sigma}^{*-1} \diag\{\mathbf{s}\}$. 

\begin{remark}\label{remark1}
Note that the existence of GKnockoff features demands the invertibility of the transformed Gram matrix $\boldsymbol{\Sigma}^{*} $. In the Supplementary Material \ref{discuss1}, we show that $\boldsymbol{\Sigma}^{*} $ is invertible  if $\mathbf{X}$ is of full column rank.
\end{remark}

The following theorem presents one of our main findings, that ${\mathbf{X}}^*$ and its GKnockoff copy $\tilde{\mathbf{X}}$  possess  the pairwise exchangeability, which is crucial to the function of GKnockoff, yet not trivial since the elements in $\mathbf{y}^*$ are no longer independent.

\begin{theorem}[\textit{pairwise exchangeability}]
Let $\mathcal{G} \subset  \mathcal{S}^c$. Then, we have
	\begin{equation*}
		[{\mathbf{X}}^{*}, \tilde{ \mathbf{X}}]_{\swap{(\mathcal{G})}}^\top {\mathbf{y}}^{*}  \stackrel{d}{=} [{\mathbf{X}}^{*}, \tilde{ \mathbf{X}}]^\top {\mathbf{y}}^{*},
	\end{equation*}
where `` $\stackrel{d}{=}$ " means equivalent in the joint distribution.
\label{lemma:exchange}
\end{theorem}
Theorem \ref{lemma:exchange} shows that the inactive features in ${\mathbf{X}}^{*}$ are pairwise exchangeable with their GKnockoff counterparts in terms of the inner product with the response variable. Under Gaussian assumption, the swapped distribution is
{
\begin{equation}
	[{\mathbf{X}}^{*}, \tilde{ \mathbf{X}}]_{\swap{(\mathcal{G})}}^\top {\mathbf{y}}^{*} \sim N([{\mathbf{X}}^{*}, \tilde{ \mathbf{X}}]_{\swap{(\mathcal{G})}}^\top {\mathbf{X}}^{*} \boldsymbol{\theta}_1, \sigma^2 [{\mathbf{X}}^{*}, \tilde{ \mathbf{X}}]_{\swap{(\mathcal{G})}}^\top \mathbf{M} [{\mathbf{X}}^{*}, \tilde{ \mathbf{X}}]_{\swap{(\mathcal{G})}} ).
	\label{df1}
\end{equation}
}
Then the pairwise exchangeability would hold only if  the expectation and covariance of swapped distribution are invariant. The invariance of expectation results from the fact that $\theta_{1j} = 0$ for $j\in\mathcal{S}^c$. The invariance of covariance, on the other hand, is a bit tricky and relies on Lemma \ref{lemma:invariant}, which states that a projection of $\tilde{ \mathbf{X}}$ is also a Gknockoff of $\mathbf{X}^*$. We refer to the Supplementary Material \ref{app:proof_invariant} for a detailed proof of Theorem \ref{lemma:exchange}.

The pairwise exchangeability motives us to extend \eqref{equ:fuse4} to an augmented regularized regression problem
\begin{equation}
	\min_{\boldsymbol\theta_1 \in \mathbb{R}^{m}, \tilde{\boldsymbol\theta}_1 \in \mathbb{R}^{m}} \frac{1}{2n}\|{\mathbf{y}}^*- {\mathbf{X}}^* \boldsymbol\theta_1 - \tilde{ \mathbf{X}} \tilde{\boldsymbol\theta}_1\|_{2}^{2}+\lambda( \| \boldsymbol\theta_1 \|_1  + \| \tilde{\boldsymbol\theta}_1 \|_1 ).
	\label{equ:aug_lasso}
\end{equation}
The regularization parameter $\lambda$ controls the sparsity level along the solution path of \eqref{equ:aug_lasso}. 

Denote $[\hat{\boldsymbol{\theta}}^\top_1(\lambda), \tilde{\boldsymbol{\theta}}^\top_1(\lambda)]^\top \in \mathbb{R}^{2m}$ the minimizer of \eqref{equ:aug_lasso}, where $\hat{\boldsymbol{\theta}}_1(\lambda)=[\hat{\theta}_{11}(\lambda),\ \ldots , \ \hat{\theta}_{1m}(\lambda)]^\top \in \mathbb{R}^m$ and $ \tilde{\boldsymbol{\theta}}_1(\lambda) =[\tilde{\theta}_{11}(\lambda), \ \ldots, \ \tilde{\theta}_{1m}(\lambda)]^\top\in \mathbb{R}^m$.
Let 
\begin{equation*}
	\lambda_{j} = \sup\{\lambda: \hat{\theta}_{1j}(\lambda) \ne 0\}, \quad \tilde{\lambda}_{j} = \sup\{\lambda: \tilde{\theta}_{1j}(\lambda) \ne 0\},
\end{equation*}
and define a vector of GKnockoff statistics $\mathbf{w}=[W_1, \ \ldots, \ W_m]^\top$ with
\begin{equation}\label{Wj}
W_j = (\lambda_{j} \vee \tilde{\lambda}_{j} ) \cdot \sign(\lambda_{j} - \tilde{\lambda}_{j}), \quad j=1,\ \ldots, \ m.
\end{equation}
A large positive value of $W_j$ provides some evidence that ${\mathbf{y}}^*$ depends on the $j$th column of ${\mathbf{X}}^*$ and hence the $j$th feature may indicate a true structural change. On the other hand, when the $j$th feature is inactive, $W_j$ should be close to 0 and is equally likely to be positive or negative. 

To control FDR at a pre-specified level $q \in [0,1]$, we follow the knockoff+ procedure \citep{Rina2015} and choose a cutoff $T(q)$ as
\begin{equation}
	T(q)=\min \left\{t \in \mathcal{W}: \frac{1 + |\left\{j: W_{j} \leq-t\right\}|}{|\left\{j: W_{j} \geq t\right\}| \vee 1} \leq q\right\},
	\label{equ:T_q}
\end{equation}
where {$\mathcal{W} = \{ |W_j|: j=1, \ \ldots, \  m \} \backslash \{ 0 \}$} and the extra term 1 in the numerator makes the choice of $T(q)$ slightly more conservative. Naturally, we estimate the active set $\mathcal{S}$ by
\begin{equation}\label{Shat}
	\hat{\mathcal{S}} = \{ j=1, \ \ldots,  \ m: W_{j}  \ge T(q) \}.
\end{equation}
Throughout this paper, we use GKnockoff filter to name the entire procedure of constructing the GKnockoff features $\tilde{\mathbf{X}}$, computing the GKnockoff statistics $\mathbf{w}$, choosing the cutoff $T(q)$, and estimating the active set by $\hat{\mathcal{S}}$. The following main theorem proves the GKnockoff filter can control FDR at any pre-specified level.
\begin{theorem}[\textit{FDR control of GKnockoff}]
	For any $q \in [0,1]$, the active set estimated by the GKnockoff filter, i.e. $\hat{\mathcal{S}}$ defined in \eqref{Shat}, satisfies
	\begin{equation}
		\mathrm{FDR}(q)=\mathbb{E}\left[\frac{ |\hat{\mathcal{S}} \cap  \mathcal{S}^c| }{| \hat{\mathcal{S}}| }\right] \le q.
	\end{equation}
	\label{th1}
\end{theorem}
\vspace{-0.5in}
Note that the construction of GKnockoff statistics is not unique and here we only exhibit one possibility as in \eqref{Wj}. See \cite{Rina2015} for more details. For instance, another appealing GKnockoff statistic is the Lasso coefficient difference (LCD) \citep{Rina2015}, that is,  $W_j = |\htheta_{1j}| - |\tilde{\theta}_{1j}|$ for $ j = 1, \ldots, m$, where $\htheta_{1j}$ and $\tilde{\theta}_{1j}$ are the solutions to \eqref{equ:aug_lasso}. Next, we study the power of the GKnockoff procedure.

\begin{theorem}[\textit{Power of GKnockoff}]
	\label{thm:powerG}
	Under Condition \ref{condition:power1}, \ref{condition:power2} and \ref{condition:min-signal} in the Supplementary Material, with probability $1-c_{\ell_1}m^{-c_{\ell_1}}$, the power of GKnockoff with LCD statistics
	$$
	\operatorname{Power} = \E\left[ \frac{|\hcalS \cap \calS|}{|\calS|} \right] \ge 1-  \frac{2C_{\ell}}{ \kappa_n },
	$$ 
	where $C_{\ell}$ and $c_{\ell_1}$ are two positive constants, and $\kappa_n \to \infty$ as $n \to \infty$.
\end{theorem}
Theorem \ref{thm:powerG} states the power of the GKnockoffs converges to 1 as $n$ goes to infinity since $\kappa_n$ goes to infinity. The proof of Theorem \ref{thm:powerG}, which is inspired by  \cite{fan2019rank}, is presented in the Supplementary Material \ref{app:powerG}. The technical challenges compared with \cite{fan2019rank} mainly lie in that the transformed error term is correlated and the design matrix is treated as fixed.

\subsection{Extended GKnockoff filter when $m < n < 2m$}
When $m < n < 2m$, we can no longer compute the GKnockoff features $\tilde{\mathbf{X}}$ from \eqref{eq:GK_compute} since it is beyond hope to find a subspace of dimension $m$ that is orthogonal to $\mathbf{X}^*$, and hence neither $\tilde{\mathbf{U}}$. To address this issue, we create $2m-n$ dummy observations and extend \eqref{equ:y_bar_dis} to the following augmented probability model
\begin{equation*}
	\left[\begin{array}{l}
		{\mathbf{y}}^{*} \\
		{\mathbf{y}}^{*}_a
	\end{array}\right]\sim N\left(\left[\begin{array}{l}
		{\mathbf{X}}^* \\
		\mathbf{0}_{(2m-n)\times m}
	\end{array}\right] \boldsymbol{\theta}_1, \sigma^{2}
	\begin{bmatrix}
		\mathbf{M} , &\mathbf{0} \\
		\mathbf{0}, &\mathbf{I}_{(2m-n)}
	\end{bmatrix} \right).
\end{equation*}
To distinguish with the GKnockoff filter introduced above, we name the GKnockoff filter based on this row-augmented data as the Extended Generalized Knockoff (EGKnockoff) filter. Theorem \ref{th2} proves that the EGKnockoff filter can also control FDR at any pre-specified level.

\begin{theorem}[\textit{FDR control of EGKnockoff}]
Denote $\hat{\mathcal{S}}_{E}:=\hat{\mathcal{S}}_{E}(q)$ the active set estimated by the EGKnockoff filter with any pre-specified level $q \in [0,1]$. Then we have
	\begin{equation}
	\mathrm{FDR}_E(q)=\mathbb{E}\left[\frac{ |\hat{\mathcal{S}}_E \cap  \mathcal{S}^c| }{| \hat{\mathcal{S}}_E| }\right] \le q.
\end{equation}
	\label{th2}
\end{theorem}
\vspace{-0.4in}

The proof of Theorem \ref{th2} is presented in the Supplementary Material \ref{app:thm2}. The EGKnockoff filter requires the sample size $n$ to be larger than the number of features $m$ since we need to estimate the unknown parameter $\sigma$ from the sample. In the next section, we propose a two-step procedure to address the high-dimensional (i.e. $m\geq n$) structural change detection and FDR control problem. We remark that in order to preserve exchangeability of EGKnockoff, the pseudo data ought to be generated from the normal distribution with mean zero and variance $\sigma^2$. See the proof of Theorem \ref{th2} for more details. As a result, the EGKnockoff can be applied as long as $\sigma^2$ is known or can be well-estimated from the data \citep{Rina2015, barber2020robust}. To avoid the estimated variance to depend on $\mathbf{y}^*$, one could adopt the data splitting strategy - one half of data to estimate the noise variance and the other to construct GKnockoff. In addition, we study the power of the EGKnockoff in Theorem \ref{thm:powerEG}. 

\begin{theorem}[\textit{Power of EGKnockoff}]
	\label{thm:powerEG}
	Under Condition \ref{condition:power1}, \ref{condition:power2} and \ref{condition:min-signal} in the Supplementary Material, with probability $1-c_{\ell_2}m^{-c_{\ell_2}}$, the power of EGKnockoff with LCD statistics
	$$
	\operatorname{Power} = \E\left[ \frac{|\hcalS \cap \calS|}{|\calS|} \right] \ge 1-  \frac{2C^\prime_{\ell}}{ \kappa_n },
	$$ 
	where $C_{\ell}^\prime $ and $c_{\ell_2}$ are two positive constants, and $\kappa_n \to \infty$ as $n \to \infty$.
\end{theorem}
Theorem  \ref{thm:powerEG} indicates  that adding pseudo data does not affect the power of EGKnockoff asymptotically. The proof of Theorem \ref{thm:powerEG} is provided in the Supplementary Material \ref{app:powerEG}.

\section{High-dimensional structural change detection}
\label{sec:high}

The GKnockoff and EGKnockoff filters require $n > m$ and hence are not applicable to high-dimensional scenarios where $n \le m$. In this section, we study high-dimensional structural change detection with FDR control and propose a two-stage procedure. We first implement a screening method to filter out a substantial number of locations where the structural changes are unlikely to exist. Then, we apply GKnockoff to the low-dimensional screened data.

\subsection{Fused sure independence screening}
\label{sec:fusion}

In this subsection, we use the piecewise constant coefficients profile model ({Scenario 1} in Section \ref{sec:frame}) as a showcase example to introduce a screening strategy for high-dimensional structural change detection problems. Recall that in this scenario, we assume the $p$ coefficients $\beta_1,\ldots,\beta_p$ can be segmented into $J+1$ groups and $\mathcal{S}=\{\tau_1,\ldots,\tau_J\}$ is the active set of all structural change locations.
Denote $X_{j}$ as the standardized  $j$th column of $\mathbf{X}$ and $\hat{\gamma}_j=X_j^\top\mathbf{y}$. Then, we define a fused screening statistic to quantify the structural change before and after a location by incorporating the information in a small neighborhood, i.e.
\begin{equation}\label{Dhat}
		\hat{\D}(j, h) = \frac{1}{h} \sum_{i = 1}^{h} \left| \hat{\gamma}_{j-i+1} - \hat{\gamma}_{j+i} \right|, \quad j=h, \ \ldots, \ p-h,
\end{equation}
where $h>0$ is a bandwidth parameter. We would expect $\hat{\D}(j, h)$ to be large if $j\in\mathcal{S}$ and  $\hat{\D}(j, h)$ to be small if there is no structural change within $\{j-h+1, \ \ldots, \ j+h\}$.

We propose to screen out the locations whose fused screening statistics are small. For a pre-specified threshold $\vartheta>0$, we can select a screened set as
\begin{equation*}
	\hat{\mathcal{A}}(\vartheta) = \{ j=1,\  \ldots, \ p-1: \hat{\D}(j, h) \ge \vartheta  \}.
\end{equation*}
The screening procedure is thereby named Fused Sure Independence Screening (FuSIS). Next, we show FuSIS enjoys a sure screening property under mild conditions, which means $\hat{\mathcal{A}}$ contains all structural changes with a probability approaching 1. To start with, we define the population fused screening statistic $\D(j, h) = \frac{1}{h} \sum_{i = 1}^{h} \left| \gamma_{j-i+1} - \gamma_{j+i} \right|$, and introduce two conditions.

\begin{condition}
 $\min_{j \in \mathcal{S }} \D(j, h) \ge 2 c_3 n^{-\kappa}$ for some constants $c_3 >0$ and $ 0 \le \kappa \le 1/2$.
\label{condition1}
\end{condition}

\begin{condition}
All structural change locations lie in $\{h, \ \ldots,\ p-h\}$.		\label{condition2}
\end{condition}

Condition \ref{condition1} is a widely used minimum signal strength condition in screening literature \citep[e.g.][]{Liu2020}. This condition is mild since it allows the minimum signal strength slowly decays to 0 as the sample size diverges.  In {the Supplementary Material \ref{app:minimum_signal}}, we provide its sufficient conditions.
Condition \ref{condition2} assumes the change points should not lie too close to the boundaries, which is common for change point detection \citep[e.g.][]{niu2012screening}. In practice, Condition \ref{condition2} can be satisfied by considering the observations near the boundaries as ``burn-in" and ``burn-out" samples where we do not detect structural changes.

	\begin{theorem}[Sure screening property]
		Under Conditions \ref{condition1} and \ref{condition2}, let $\vartheta \le \min_{j \in \mathcal{S}}\D(j, h)/2 $, we have
		\begin{equation}
			\operatorname{Pr}(\mathcal{S} \subset \widehat{\mathcal{A}}(\vartheta)) \geq 1-O\left(hJ \exp \left\{-c_{4} n^{1-2 \kappa}\right\}\right),
		\end{equation}
		where $c_4>0$ is a positive constant and $J = |\mathcal{S}|$.
		\label{thm2}
	\end{theorem}
 The proof of Theorem \ref{thm2} is given in the Supplementary Material\ref{sec:sis}.

\subsection{Bandwidth selection}
\label{sec:multiband}

The  bandwidth parameter $h$ plays an essential role in FuSIS. Next, we introduce a  data-driven bandwidth selection procedure. Let $h_1,\ \ldots, \ h_B$ be a sequence of grid points. For a given grid point $h_k$, $k=1, \ \ldots, \ B$, denote $\hat{\mathcal{A}}_k(\vartheta)=\{\hat{\tau}_{k1},\ldots,\hat{\tau}_{k\hat{J}_k}\}$ the set screened by FuSIS with the bandwidth $h_k$, where  $\hat{J}_k=|\hat{\mathcal{A}}_k(\vartheta)|$. The set $\hat{\mathcal{A}}_k(\vartheta)$ naturally divide the features in $\mathbf{X}$ into $\hat{J}_k+1$ homogeneous groups, say $\hat{G}_1,\ \ldots, \ \hat{G}_{\hat{J}_k+1}$, such that the coefficients share the same value within each group.
For each $\hat{\mathcal{A}}_k(\vartheta)$, we can solve a constrained ordinary least squaresl problem
\begin{align}
    	\min_{\mathbf{b}\in\mathbb{R}^{p}} \|\mathbf{y} - \mathbf{X} \mathbf{b}\|^2_2 \ 
    	\text{ subject to } \  b_{1} = \cdots  =b_{\hat{\tau}_{k1} }; \  \ldots; \  b_{\hat{\tau}_{k\hat{J}_k}+1} = \cdots =b_{p}. \nonumber
\end{align}
This optimization problem is equivalent to
\begin{equation}
	\min_{\boldsymbol\nu \in \mathbb{R}^{\hat{J}_{k}+1}} || \mathbf{y}-\mathbf{X}\mathbf{Q}_k\boldsymbol{\nu} ||_2^2,
	\label{ols1}
\end{equation}
where $\mathbf{Q}_k$ is a $p \times (\hat{J}_{k}+1)$ matrix, whose $(i,j)$th entry equals to $1$ if the $i$th feature in $\mathbf{X}$ belongs to $\hat{G}_j$ and $0$ otherwise. The solution of \eqref{ols1} admits a closed form
\begin{equation*}
\hat{\boldsymbol{\nu}}_k= \left\{(\mathbf{XQ}_k)^\top(\mathbf{XQ}_k)\right\}^{-1} (\mathbf{XQ}_k)^\top \mathbf{y}.
\end{equation*}
Further, we can define the $R^2$ associated with $\hat{\boldsymbol{\nu}}_k$, and hence $h_k$,  as $R^2_k$.  The empirical optimal bandwidth is defined as
\begin{equation}\label{h_opt}
h_{opt} (\vartheta)= \arg\max_{h_k\in\{h_1,\ldots,h_B\}} R^2_k,
\end{equation}
and the resulting screened set is dentoed as $\hat{\mathcal{A}}_{opt}(\vartheta)$.
We summarize the entire FuSIS procedure with bandwidth selection in Algorithm \ref{algorithm1}.

  \begin{algorithm}
 	\caption{FuSIS with bandwidth selection}
 	\begin{algorithmic}[1]
 		\State \textbf{Input:} Observed data $(\mathbf{X},\mathbf{y})$, bandwidth grid points $h_1, \ldots, h_B$, and a threshold $\vartheta$. 
 		\State \textbf{FuSIS:} For  $k=1,\ldots, B$, apply FuSIS to $(\mathbf{X},\mathbf{y})$ with bandwidth $h_k$. Obtain the $k$th screened set $\hat{\mathcal{A}}_k(\vartheta)$ and the associated $R^2_k$.
 		\State \textbf{Bandwidth selection:}
 		Define the optimal bandwidth as (\ref{h_opt}).
		 		\State \textbf{Output:} $\hat{\mathcal{A}}_{opt}(\vartheta)$.
 	\end{algorithmic}
 	\label{algorithm1}
 \end{algorithm}

\subsection{High-dimensional generalized knockoff}

In this subsection, we propose a two-stage procedure named High-dimensional Generalized Knockoff filter (HGKnockoff filter) to detect structural changes in high-dimensional scenarios and  control FDR at a pre-specified level. To avoid the mathematical and empirical challenges cased by reusing the data, we adopt a data splitting strategy for the two steps. To be specific, We randomly partition $(\mathbf{X},\mathbf{y})$ into two subsamples $(\mathbf{X}^{(1)},\mathbf{y}^{(1)})$ and $(\mathbf{X}^{(2)},\mathbf{y}^{(2)})$ with sample sizes $n_1$ and $n_2 = n - n_1$, respectively.

The two stages of the HGKnockoff filter are introduced as follows:
\begin{itemize}
\item[(1)] \textsc{FuSIS stage:} Apply Algorithm \ref{algorithm1} to $(\mathbf{X}^{(1)}, \mathbf{y}^{(1)})$ with a threshold $\vartheta$ such that the screened set  $\hat{\mathcal{A}}_{opt}(\vartheta)$ contains less than $n_2/2$ elements, i.e. $|\hat{\mathcal{A}}_{opt}(\vartheta)|<n_2/2$.

\item[(2)] \textsc{GKnockoff stage:} Denote $\mathbf{X}^{(2)}_{\mathrm{FuSIS}}$ the sub-matrix of $\mathbf{X}^{(2)}$ whose column corresponding to $\hat{\mathcal{A}}_{opt}(\vartheta)$. Then, we apply the GKnockoff filter to $(\mathbf{X}^{(2)}_{\mathrm{FuSIS}}, \mathbf{y}^{(2)})$ to detect structural changes while controlling FDR at a pre-specified level $q$. The final estimator of the active set is denoted as $\hat{\mathcal{S}}_H:=\hat{\mathcal{S}}_H(\vartheta, q)$.
\end{itemize}

In Theorem \ref{thm:fdr} below, under mild conditions, we prove the HGKnockoff filter can control FDR at any pre-specified $q \in [0, 1]$.

\begin{theorem}
	(a) Under Conditions \ref{condition1} and \ref{condition2}, for any $q\in[0,1]$, the HGKnockoff filter satisfies
	\begin{equation}\label{asymFDR}
	\lim_{n\to\infty}\mathrm{FDR}_H (q)=\mathbb{E}\left[\frac{| \mathcal{S}^c \cap \hat{\mathcal{S}}_H|}{|\hat{\mathcal{S}}_H| }\right] \le q.
\end{equation}

(b) Furthermore, conditional on the sure screening event $\mathcal{E} = \{ \mathcal{S} \subset \hat{\mathcal{A}}_{opt}(\vartheta)\}$, we can get a finite sample guarantee of FDR control
	\begin{equation}
	\mathrm{FDR}_H(q)=\mathbb{E}\left[\frac{| \mathcal{S}^c \cap \hat{\mathcal{S}}_H|}{|\hat{\mathcal{S}}_H| } \Big| \mathcal{E} \right] \le q.
\end{equation}
	\label{thm:fdr}
\end{theorem}
\vspace{-0.4in}

\section{Simulation Studies}
\label{sec:simulation}
In this section, we simulate various structural change detection experiments to evaluate the empirical performance of GKnockoff, FuSIS, and HGKnockoff.  We also compare the proposed methods with some popular competitors in the literature.

\subsection{Simulations for the GKnockoff filter}
\label{GKnockoffSimulation}
We apply the GKnockoff filter to study the two structural change detection scenarios discussed in Section \ref{prob state}. For the B-Y procedure, we first estimate the regression coefficients $\hbbeta$ and noise variance $\hsigma^2$, then for testing the hypothesis $H_{0j} : \bd_j^\top \bbeta = 0$, $j=1,\ldots,m$, compute the corresponding p-value $p_j$ through the t-statistic $t_j = \bd^\top \hbbeta/\hat{\sigma}^*_{j} $ where $\hat{\sigma}^*_{j} = \sqrt{\hsigma^2 \bd_j^\top (\bX^\top \bX)^{-1} \bd_j}$. Then the standard B-Y procedure is applied to obtain the selected set. For the permutation procedure, we randomly permute rows of design matrix so that the permuted predictors no longer possess predictive effect on the response - thus can be treated as ``knockoffs" to some extent. We also apply the B-Y method \citep{Benjamini2001} and the classical permutation-based method to these scenarios for comparison purpose. We will discuss the permutation-based method at the end of this subsection, and show that it fails to control FDR in our simulations settings.  Therefore, we focus on comparing  the GKnockoff filter and the B-Y method in terms of the estimated FDR  and the empirical power. Throughout this subsection, we set the error variance $\sigma^2=1$,  $\mathbf{X} = (\mathbf{x}_1, \ldots, \mathbf{x}_n)^\top$, and draw $\mathbf{x}_i$'s independently from $N(\mathbf{0}, \mathbf{\Sigma}_{p \times p})$ where $ \boldsymbol{\Sigma}_{(i,j)}  = \rho^{|i-j|}$ for some $\rho \in [0,1)$. The nominal FDR level is fixed to be $q=0.2$.
The active set of structural change locations is set to be $\mathcal{S}=\{\tau_1, \ \ldots, \ \tau_J \}$. For each case, we simulate 200 replications. To be specific, the estimated FDR and the empirical Power are defined by
\begin{equation*}
    \widehat{\text{FDR}} = \frac{1}{200} \sum_{i=1}^{200} \frac{|\hat{\mathcal{S}}_i\cap \mathcal{S}^c|}{|\hat{\mathcal{S}}|}
    \quad \text{and} \quad
    \widehat{\text{Power}} =  \frac{1}{200} \sum_{i=1}^{200} \frac{|\hat{\mathcal{S}}_i \cap \mathcal{S} |}{ |\mathcal{S} |},
\end{equation*}
where $\hat{\mathcal{S}}_i$ is the estimated active set in the $i$th replication.

\noindent{\bf Experiment 4.1: Piecewise constant coefficients profile}

 Consider the piecewise constant coefficients profile model in {Scenario 1}. We set $n = 350$ and $p= 100$. The true coefficients are set to be
\begin{equation*}
    \beta_{\tau_{k-1}+1}= \ \cdots \  =\beta_{\tau_k}=(-1)^k A, \quad \text{for} \quad  k=1, \ \ldots, \ J,
\end{equation*}
where $A$ is a positive parameter that controls the signal amplitude and we denote $\tau_0=0$.  We choose $J$, $A$ and the $\rho$ as follows.
	\begin{itemize}
		\item [(1)] Fix $A = 0.12$, $\rho = 0$, and let $J$ vary in $\{9, 10, 11, 12, 13,14\}$.
		\item [(2)] Fix $J = 13$, $\rho = 0$, and let $A$ vary in  $\{0.09, 0.10, 0.11, 0.12,0.13,0.14\}$.
		\item [(3)] Fix $J = 13$, $A = 0.12$. and let $\rho$ vary in $\{0, 0.06, 0.12, 0.18, 0.24, 0.30\}$.
	\end{itemize}

\begin{figure}[H]
		\centering
		\includegraphics[width=7.2cm, height=6cm]{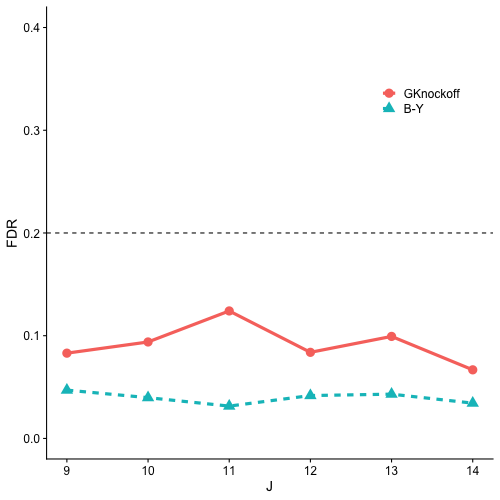}
		\includegraphics[width=7.2cm, height=6cm]{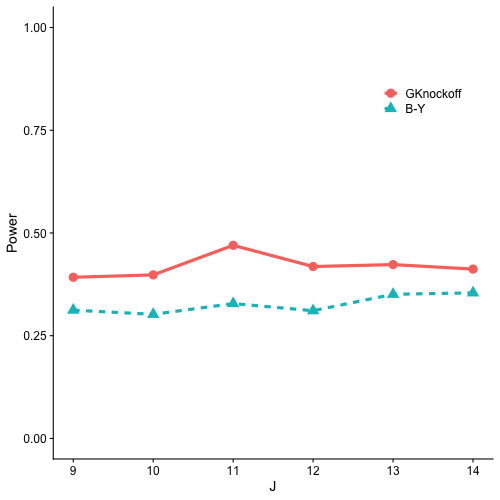}
		\caption{FDR and power with respect to $J$ for GKnockoff and B-Y in Experiment 4.1. \label{fig:lm_vary_k}}
	\end{figure}

\begin{figure}[H]
		\centering
		\includegraphics[width=7.2cm, height=6cm]{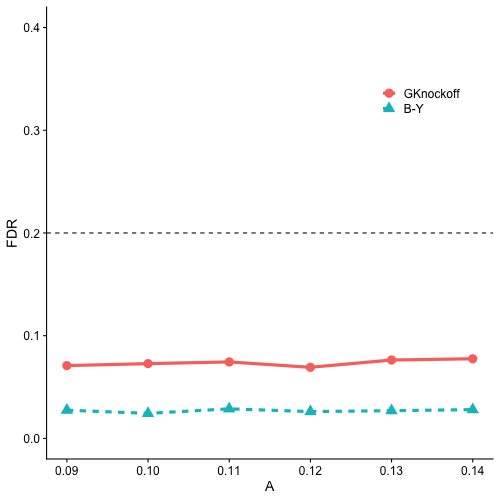}
		\includegraphics[width=7.2cm, height=6cm]{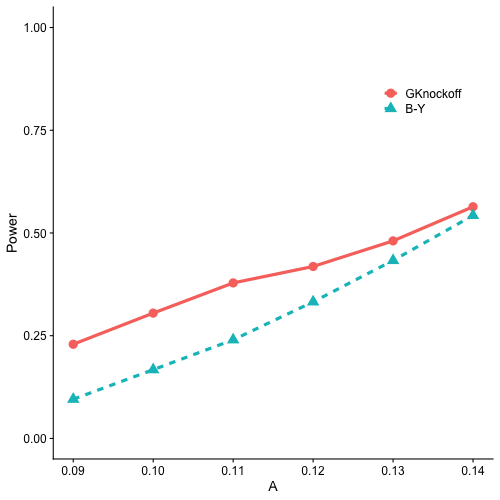}
		\caption{FDR and power with respect to $A$ for GKnockoff and B-Y in Experiment 4.1. \label{fig:lm_vary_A}}
	\end{figure}

\begin{figure}[H]
		\centering
		\includegraphics[width=7.2cm, height=6cm]{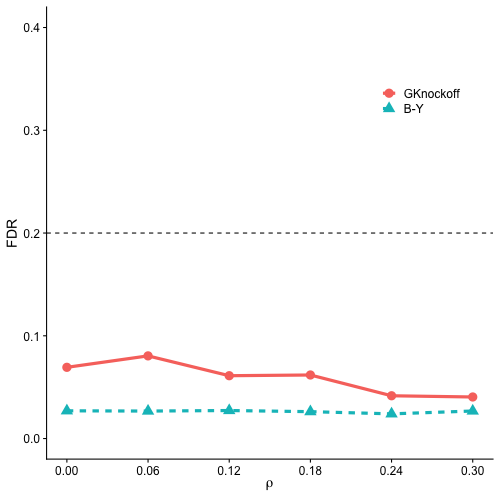}
		\includegraphics[width=7.2cm, height=6cm]{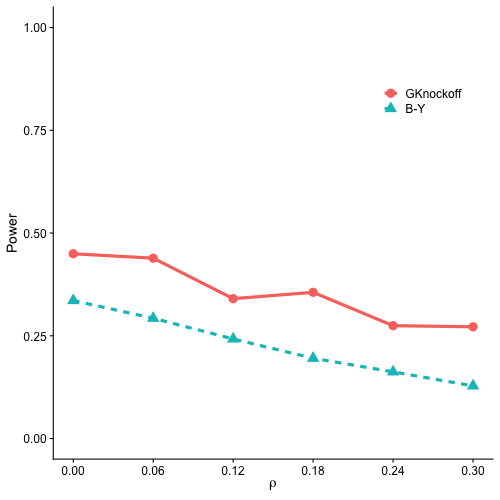}
		\caption{FDR and power with respect to $\rho$ for GKnockoff and B-Y in Experiment 4.1.
		\label{fig:lm_vary_rho}}
	\end{figure}

The simulation results are summarized in  Figures \ref{fig:lm_vary_k}, \ref{fig:lm_vary_A} and \ref{fig:lm_vary_rho}. {Figure \ref{fig:lm_vary_k} summarizes the estimated FDR and the empirical power with a fixed $A$ and an increasing $J$. We observe that both methods can control FDR under the pre-specified level. The GKnockoff filter has higher empirical powers than the B-Y method in all cases. Figure \ref{fig:lm_vary_A} summarizes the estimated FDR and the empirical power with a fixed $J$ and an increasing $A$. Again, both methods can control FDR under the pre-specified level and the GKnockoff filter outperforms the B-Y method in terms of empirical powers. \ref{fig:lm_vary_rho} shows similar phenomenon as the previous experiment, in which both methods can control FDR at the pre-specified level $q = 0.2$; regarding the empirical power, the GKnockoff filter uniformly outperforms the B-Y method.}

\noindent{\bf Experiment 4.2: Integration analysis from multiple data sources}

Consider the integration analysis from $K$ data sources as discussed in Scenario 2. We set $p=40$. The sample size of the $k$th source, i.e. $n^{(k)}$, is independently drawn  from $Poission(\zeta)$, where $\zeta = 100$, for $k=1, \ \ldots, \ K$. The number of structural changes $J$ now stands for the total number of distinct coefficients from adjacent data sources. If $(k, \tau_j)$ is a change position, we set $\beta_{\tau_j}^{(k)} = - \beta_{\tau_j}^{(k+1)}$. The amplitude $A$ is accordingly defined as $A=|\beta_{j}^{(k)}|$ for all $k=1, \ \ldots, \ K$ and $j=1, \ \ldots, \ p$. The $i$th sample from the $k$th source $\mathbf{x}_i^{(k)}$ is independently generated from  $N(0, \mathbf{I}_p), i= 1, \ \ldots, \ n_k$ and $k=1, \ \ldots, \ K$.
Further, we choose $K$, $A$ and $J$ as follows. 	
\begin{itemize}
		\item [(1)] Fix $K = 5$, $A = 0.25$, and let $J$ vary in $\{15, 17, 19, 21, 23, 25\}$.
		\item [(2)] Fix $J = 20$, $A = 0.25$, and let $K$ vary in $\{3, 4, 5, 6, 7, 8\}$.
	\end{itemize}

\begin{figure}[H]
		\centering
		\includegraphics[width=7.2cm, height=6cm]{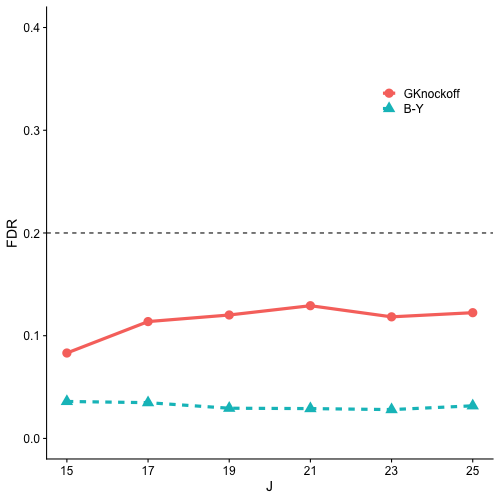}
		\includegraphics[width=7.2cm, height=6cm]{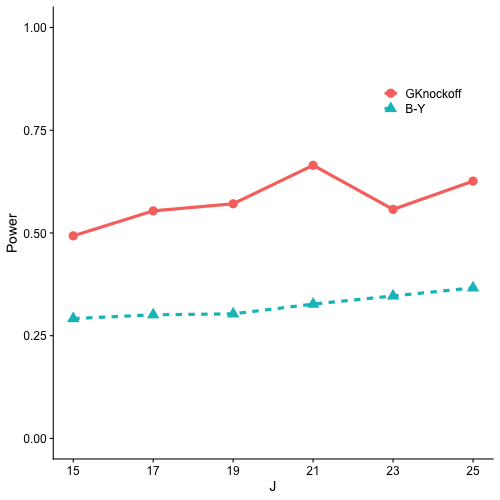}
		\caption{FDR and power with respect to $J$ for GKnockoff and B-Y in in Experiment 4.2.} \label{fig:inte_vary_k}
	\end{figure}

\begin{figure}[H]
		\centering
		\includegraphics[width=7.2cm, height=6cm]{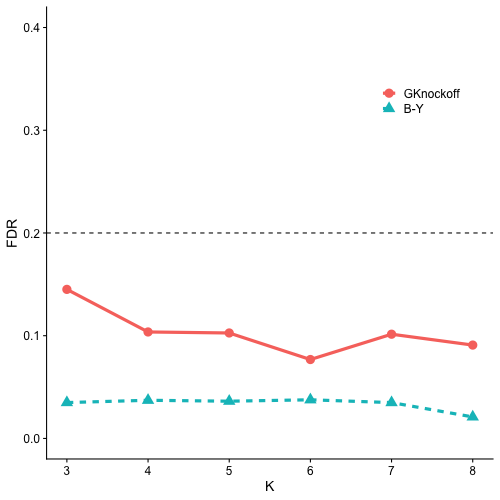}
		\includegraphics[width=7.2cm, height=6cm]{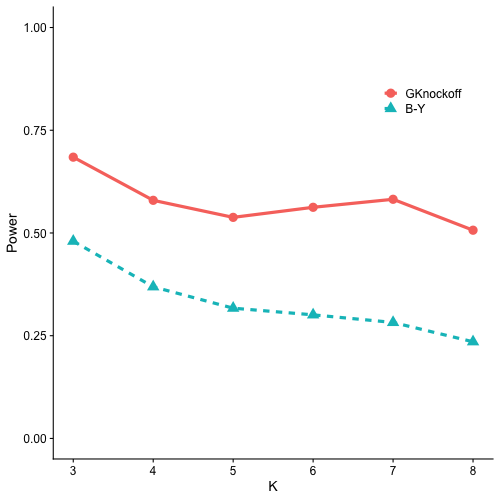}
		\caption{FDR and power with respect to $K$ for GKnockoff and B-Y in in Experiment 4.2.}  \label{fig:inte_vary_source}
	\end{figure}

The simulation results are summarized in Figures \ref{fig:inte_vary_k} and \ref{fig:inte_vary_source}. Again, both methods can successfully control FDR at the pre-specified level $q = 0.2$, and the GKnockoff filter gains significantly more empirical power than the B-Y method.
	
Next, we compare the FDR control performance of the Gknockoff filter with a permutation-based method in all two experiments studied above. The permutation-method constructs ``knockoff" of $\mathbf{X}^*$ by randomly permuting its rows.  For Experiment 4.1,  we set $n = 350$,  $p=100$, $A=0.1$, $J = 10$ and $\rho=0$. For Experiment 4.2, we set $\zeta = 100$, $p=40$, $A = 0.25$, $K= 5$, $J = 20$ and $\rho=0$. Table \ref{tab:1} reports the estimated FDR of the two methods, which shows the permutation-based method fails to control FDR at $q=0.2$ for all two structural change problems. We argue that the permutation-based method, though straightforward, can not address the dependence in the noise $\mathbf{M}\boldsymbol{\epsilon}$ and hence does not enjoy the pairwise exchangeability.

	\begin{table}[H]
		\caption{Estimated FDR by permutation and GKnockoff}
		\centering
		\label{tab:1}
		\begin{tabular}{ccc}\hline\hline
			 & piecewise constant coefficients & Integration analysis\\ \hline
			Permutation          & 0.611         & 0.595        \\
			GKnockoff         & 0.096         & 0.122       \\
			\hline\hline
		\end{tabular}
	\end{table}

\subsection{Simulations for FuSIS}
\label{FuSISsimulation}

In this subsection, we use simulated experiments to assess the finite sample performance of FuSIS for screening coefficient changes in Experiment 4.2. Within each replication, we compute and rank $\hat{\D}(h,j) $ in a descending order and choose the first $n-1$ locations as the selection set $\hat{\mathcal{A}}$. We set sample size and dimensionality  to be  $(n,p)=(300,\  1000)$ and $(n, p)= (1500, \ 10000)$.  The active set of structural change locations is set to be $\mathcal{S}=\left\{\frac p{10}, \frac {2p}{10},\ \ldots, \ \frac{9p}{10}\right\}$ with $J=9$, and the signal amplitude $A=0.1$.
The covariates, i.e. $\mathbf{x}_i$'s, are independently drawn from $ {N}(\mathbf{0}, \boldsymbol{\Sigma})$, where $\boldsymbol\Sigma$ admits one of the following two forms.
	\begin{itemize}	
		\item [(1)] (\textit{AR structure}) $ \boldsymbol{\Sigma}_{(i,j)}= \rho^{|i - j|}, \  i,j=1, \ \ldots,\  p.$
		\item [(2)] (\textit{Group structure}) $\boldsymbol\Sigma$ is a block diagonal matrix, with ten $\frac p{10}\times \frac p{10}$ dimensional matrices on the diagonal, each of which is defined as $\boldsymbol\Sigma^{(0)}$, where  $\boldsymbol\Sigma^{(0)}_{(i,j)}= \rho^{|i-j|}, \ i, j=1, \ \ldots, \ \frac p{10}$.
	\end{itemize}
We take $\rho=0.3, \ 0.6$ and $0.9$, respectively. Based on 1000 simulation replications, we assess the sure screening property of FuSIS via the coverage proportion of all true coefficient change locations. For bandwidth selection, we demonstrate two methods: (a) a fixed bandwidth that is chosen such that at most one change occurs within each $h$-neighborhood; specifically, $h=25$ when $p = 1000$, and $h=100$ when $p=10000$; (b) an optimal bandwidth selected by the data-driven bandwidth method introduced in Section \ref{sec:multiband}. The results are reported in Table \ref{tab:screen}, from which one can see that the coverage proportions are all close to 1. In addition, the optimal bandwidth generally yields a larger coverage rate than the fixed bandwidth. 	
\begin{table}[H]
		\caption{Coverage proportion of FuSIS}
		\label{tab:screen}
		\centering
		\begin{tabular}{ccccccccccccc}
			\hline
			\hline
			&& \multicolumn{2}{c}{$p=1000$, $n=300$} & &\multicolumn{2}{c}{$p=10000$, $n=1500$} \\ \cline{3-4}\cline{6-7}
			&$\rho$         & Fixed $h$     & Optimal $h$ &   & Fixed  $h$  & Optimal $h$ \\
			\hline
		                 &      0.3    & 0.886     & 0.918   &  & 0.918     & 0.942    \\
	AR structure	&	0.6   & 0.958       & 0.964      &  & 0.964     & 0.983    \\
	                         &      0.9     & 0.966    & 0.983  &   & 0.971    & 1       \\ \hline
	        		        &      0.3    & 0.876     & 0.927 &    & 0.917    & 0.974   \\
Group structure        &      0.6   & 0.933        & 0.966 &       & 0.966    & 1   \\
	                        &       0.9     & 0.972    & 0.982 &  & 0.982    & 1     \\ \hline\hline
		\end{tabular}
	\end{table}

\subsection{Simulations for the HGKnockoff filter}
\label{sim:HGK}

In this subsection, we access the performance of the HGKnockoff filter for a high-dimensional piecewise constant coefficients profile model. Note that the B-Y method is not applicable when $p>n$, and hence we adopt the same data splitting technique to first screen the potential structural changes and then apply the B-Y method to the screened features. We name this method the screened B-Y method. In addition, we also consider the sequential B-H method \citep{g2016sequential} as a competitor.

\begin{figure}[H]
		\centering
		\includegraphics[width=7.2cm, height=6cm]{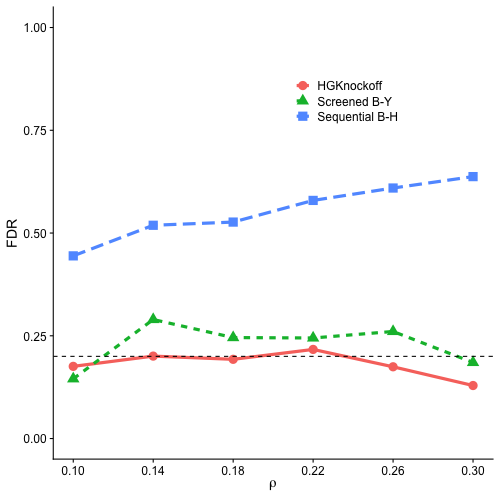}
		\includegraphics[width=7.2cm, height=6cm]{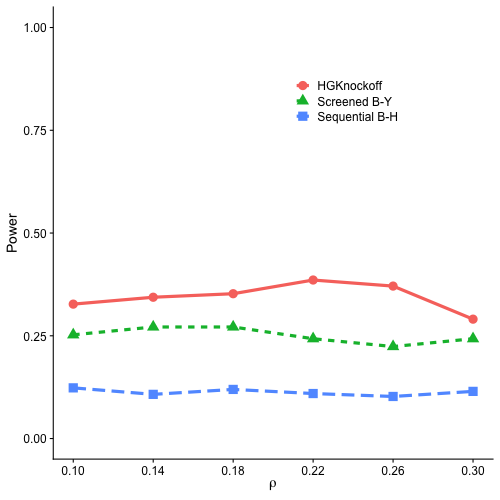}
		\caption{FDR and power trend of HGKnockoff, screened B-Y and sequential B-H with respect to $\rho$ for high-dimensional piecewise constant coefficients profile model under AR structure. \label{fig:split_method_toeplitz}}
	\end{figure}

\begin{figure}[H]
		\centering
		\includegraphics[width=7.2cm, height=6cm]{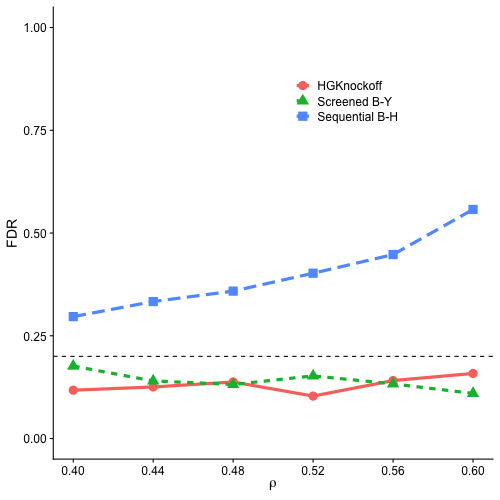}
		\includegraphics[width=7.2cm, height=6cm]{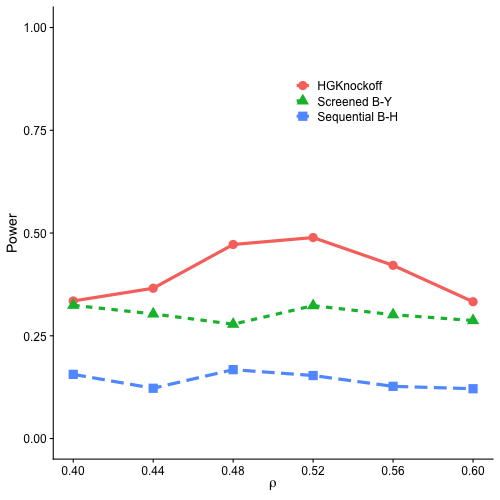}
		\caption{FDR and power trend of HGKnockoff, screened B-Y and sequential B-H with respect to $\rho$ for high-dimensional piecewise constant coefficients profile model under group structure. \label{fig:split_method_group}}
	\end{figure}

We follow a similar simulation setup as in Section \ref{FuSISsimulation} except for the following aspects. We set $n= 900$, $p= 1000$, $A = 0.15$, and $J=8$. We vary $\rho$ from  $0.1$ to $0.3$ for the AR structure, and from $0.4$ to $0.6$ for the group structure. The sample is randomly partitioned into two halves, one for FuSIS and the other one for structural change detection with FDR control.
The simulation results, measured by the estimated FDR and the empirical power, are summarized in Figures \ref{fig:split_method_toeplitz} and  \ref{fig:split_method_group}.
We observe that the sequential B-H method fails to control FDR at the pre-specified level $q=0.2$, partly due to the simulation setup violates the independence assumption. The screened B-Y method also does not control FDR well, especially for the AR structure setting. In contrast, the HGknockoff filter controls FDR at $q=0.2$. Moreover, the HGknockoff filter has the highest empirical power among the three competitors.
Notably, the power trends of HGknockoff behave like inverted-U curves, which reflect the trade-off between controlling FDR and satisfying the sure screening property.
	
\section{Real data analysis}
\label{sec:realdata}
In this section, we apply the proposed GKnockoff filter to analyze a Chinese macroeconomic dataset, aiming to explore the relationship between Gross Domestic Product (GDP) and industry structure for different provinces in China.  
The past decades have witnessed an extraordinary growth of the Chinese economy, with its GDP ranked second in the world. However, rapid economic growth also brings about uneven development across different Chinese provinces as a price. Recently, the government has turned down the voice of high-speed growth but emphasized ``high-quality growth", which emphasized the driving effect of GDP on the industrial structure, especially the secondary industry. Therefore, we are motivated to study the effect of GDP  on the proportion of the secondary
industry, which may differ among provinces; meanwhile, some provinces might perform similarly. We target to discover heterogeneous effects among provinces. \cite{zhong2021estimation} studied a similar problem, but on city level, by conducting multi-kink quantile regression. We view it from a different perspective of integration analysis in this paper, and aim to detect the coefficient changes across provinces in the multiple-source model. 

The dataset was collected from Organization for Economic Cooperation and Development database (OECD)\footnote{https://insights.ceicdata.com/}. After removing missing values and provinces with less than 3 cities, the dataset contains various economic measurements in 245 cities across 23 provinces of China in year 2016. We first sort the provinces in an descending order according to the GDP per capita, following the assumption that provinces with similar economic development should possess similar driven effects of GDP on industrial structure \citep{zhong2021estimation}. The ordered provinces by GDP per capita are Jiangsu, Inner Mongolia, Zhejiang, Shandong, Fujian, Guangdong, Hubei, Jiangxi, Jilin, Hunan, Guizhou, Ningxia, Hebei, Liaoning, Henan, Anhui, Guangxi, Sichuan, Heilongjiang, Shanxi, Shaanxi, Gansu and Yunnan.  

The response $y^{(k)}_{i}$ and exposure variable $x_{i}^{(k)}$ are respectively taken to be the proportion of secondary industry and GDP per capita of the $i$th city in the $k$th ordered province. Furthermore, as illustrated by \cite{zhong2021estimation}, fiscal expenditure (FE) and fixed assets investment (FAI) are also associated with industry structure. Therefore, we establish the following model for the $i$th city in the $k$th province as
\begin{equation}\label{realdata}
	y^{(k)}_{i} = \beta^{(k)}  x_{i}^{(k)} + \alpha_1  z_{1i}^{(k)} + \alpha_2 z_{2i}^{(k)}  + \epsilon_i^{(k)},
\end{equation}
where $\beta^{(k)}$ is the driven effect of economic growth on the secondary industry for the $k$th ordered province, $\alpha_1$ and $\alpha_2$ are homogeneous effects of FE (denoted as $z_{1i}^{(k)}$) and FAI ($z_{2i}^{(k)}$), and $\epsilon_i^{(k)}$ follows $N(0, \sigma^2)$ independently. To detect the heterogenous effects, we assume 
$$\beta^{(1)} = \ldots =  \beta^{(\tau_1)} \ne  \beta^{(\tau_1+1)} = \ldots = \beta^{(\tau_2)} \ne  \beta^{(\tau_2+1)} = \ldots = \beta^{(\tau_J)} \ne  \beta^{(\tau_J+1)} = \ldots = \beta^{(23)},$$ 
with $\calS  = \{ \tau_1, \ldots, \tau_J\}$ denoted as the index set of true coefficient changes.  

We apply the GKnockoff filter to fit model \eqref{realdata}, under a pre-specified FDR level $q=0.2$. We also conduct the B-Y procedure and the regular fused Lasso without FDR control for comparison purpose. The mean prediction errors of the three methods, as well as the obviously overfitting ordinary least squares (OLS) method, are reported in Table \ref{tab:3}, from which we observe the superior performance of GKnockoff over other methods in terms of prediction error.

\begin{table}[H]
  \caption{Prediction errors of four methods}
 \centering
  \label{tab:3}
 \begin{tabular}{ccccc}\hline\hline
  & GKnockoff & B-Y   & Fused Lasso & OLS   \\ \hline
  MPSE & 1.093    & 1.345 & 1.142       & 1.227 \\
   \hline\hline
 \end{tabular}
\end{table}

\begin{figure}[H]
	\centering
	\includegraphics[width=15cm, height = 6.5cm]{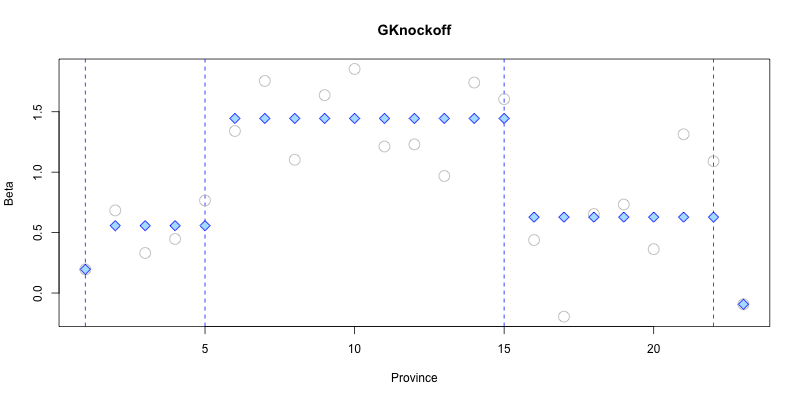}
	\includegraphics[width=15cm, height = 6.5cm]{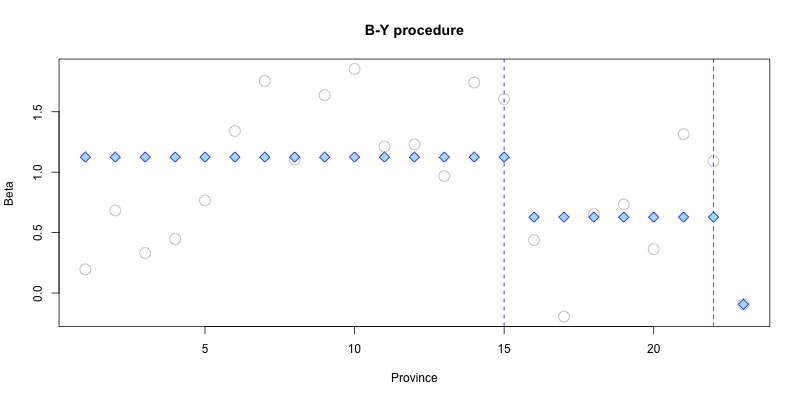}
	\includegraphics[width=15cm, height = 6.5cm]{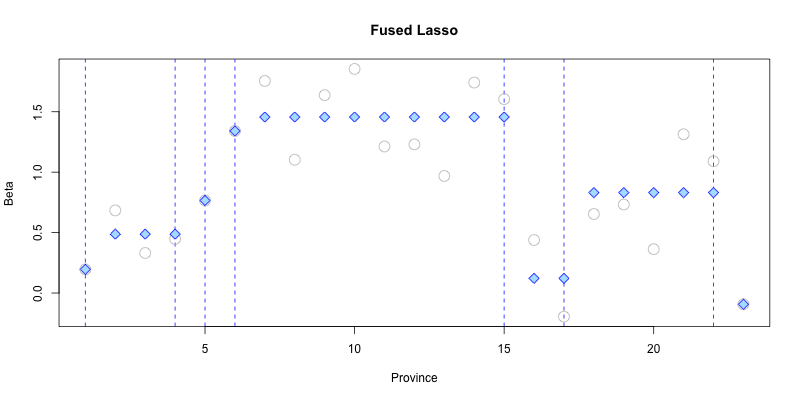}
	\caption{Effects of GDP on proportion of secondary industry estimated by GKnockoff, B-Y and Fused Lasso without FDR control, respectively. Blue diamonds represent $\hbeta^{(k)}$'s for the $k$th provinces estimated by the three methods. The change positions are represented by blue dashed lines. Grey circles are estimated coefficients fitted by ordinary least squares for each province separately. {\label{fig:realdata}}}
\end{figure}

The change positions of effects estimated are $\{1, 5, 15, 22\}$, $\{15, 22\}$ and $\{1, 4, 5, 6, 15, 17, 22\}$ by GKnockoff, B-Y and fused Lasso without FDR control, respectively, as shown in  Figure \ref{fig:realdata}. Compared with the regular fused Lasso that clearly contains many falsely discovered changes, both GKnockoff and B-Y are able to control FDR. The GKnockoff procedure implies that the effect of GDP per capita on the proportion of secondary industry follows a reversed U shape. The GDP has more driven effects for provinces with moderate economic sizes, while this effect will be diminished when the province's GDP per capita becomes larger or smaller. Meanwhile, the B-Y procedure indicates more driven effects for larger economic sizes (thus smaller rankings). Therefore, GKnockoff achieves higher detection power than B-Y in this analysis, since the phenomenon discovered from GKnockoff is more consistent with existing literature. For instance, \cite{zhu2012understanding} stated that manufacturing benefits from more production externalities than does agriculture, which means the secondary industry will grow faster than other sectors as economic size grows. However, the regional inequality may lead to different driven effects \citep{cheong2014impacts}. For more developed regions, the economy may enter the ``New Normal'' status, so that the growth of manufacturing sector might in turn slow down \citep{chen2019china}. 

\section{Conclusion}
Inspired by a structural change detection problem for the driven effects of economic development on the secondary industry, we developed a generalized knockoff procedure (GKnockoff) for selecting structural changes while controlling the false discovery rate (FDR). Upon identifying potential structural changes rather than individual features, we adopted the generalized Lasso approach via introducing some full-row-rank transformation matrix for the original coefficient vector. We carefully studied its selection consistency and asymptotic power. The transformed data used in generalized Lasso violates the independence assumption which is crucial to the theoretical guarantees of the classical knockoff. Seeing this, we proposed to construct knockoffs based on the projected design matrix, that accommodates the dependence structure of transformed data. We established the pairwise exchangeability of the GKnockoff design and proved its capability to rigorously control FDR under finite samples. For high-dimensional features, we proposed a new screening technique, called FuSIS, which is of its own significance, that reduces dimensionality by filtering out redundant structural changes. Further, we adopted a data splitting technique, named high-dimensional GKnockoff (HGKnockoff), to first reduce dimensionality and then apply GKnockoff respectively on two halves of data. The sure screening property of FuSIS and the capability of HGKnockoff to control FDR were also proved. We showed the powers of GKnockoff and EGKnockoff approach to one under mild conditions. Simulation studies empirically verified the outstanding performance of GKnockoff and HGKnockoff in terms of FDR control and power, as well as the sure screening property of FuSIS. We applied the proposed method to analyze a macroeconomic dataset that describes the structural changes of driven effects of GDP on the secondary industry. It turns out that the GKnockoff filter yields a higher power compared with the B-Y procedure.

\newpage

\appendix

\makeatletter
\renewcommand \thesection{S\@arabic\c@section}
\renewcommand\thetable{S\@arabic\c@table}
\renewcommand \thefigure{S\@arabic\c@figure}
\makeatother

\renewcommand{\thecondition}{S.\arabic{condition}}
\renewcommand{\thesubsection}{S.\arabic{subsection}}
\renewcommand{\thelemma}{S.\arabic{lemma}}
\renewcommand{\thetheorem}{S.\arabic{theorem}}
\renewcommand{\theequation}{S.\arabic{equation}}
\renewcommand{\thetable}{S.\arabic{table}}
\renewcommand{\thefigure}{S.\arabic{figure}}

\section{Supplementary Material}

\subsection{Proofs of Theorems \ref{the:power}}
The proof of Theorem \ref{the:power} follows Theorem 1 in \cite{Wainwright2009}, by realizing that the error term in the transformed model is normally distributed and the spectral norm of the projection matrix is upper bounded by 1.

{
	\subsection{Discussion of Remark \ref{remark1}}
	\label{discuss1}
	In this section, we show that although $\mathbf{M}$ is not of full rank, $\mathbf{X}^*$ is of full column rank if only $\mathbf{X}$ is. Since the augmented matrix $\tilde{\mathbf{D}} \in \mathbb{R}^{p \times p} $ is full rank, the rank of $\bX \tilde{\mathbf{D}}^{-1} = \bX[ \bZ_{p \times m}, \mathbf{F}_{p \times (p -m)}]$ is $p$ provided that $\mathbf{X}$ is of full column rank and $n>p$. This implies $\bX \bZ$ is of full column rank with rank $m$, and the columns of $\bX \bZ$ and the colums of $\bX \mathbf{F}$ are linearly independent. Furthermore, recall that the projection matrix $\bM=\mathbf{I}_n - (\mathbf{X}\mathbf{F}) [(\mathbf{X}\mathbf{F})^\top (\mathbf{X}\mathbf{F})]^{-1} (\mathbf{X}\mathbf{F})^\top$. {Thus the intersection of the kernel space of $\bf{M}$ and the space spanned by the columns of $\bX \bZ$ only includes $\boldsymbol{0}$}. Therefore, by Proposition 4.2.7 of \cite{rao1998matrix}, $\Rank(\bX^*)$ = $\Rank(\bM \bX \bZ) = \Rank(\bX \bZ) = m$, which implies $\mathbf{X}^*$ is full column rank. In a word, the rank deficit of projection matrix $\mathbf{M}$ does not affect the construction of GKnockoff. 
}

\subsection{Proof of Theorem \ref{lemma:exchange}}
\label{app:proof_invariant}
We first establish the following lemma as preparation for the pairwise exchangeability of GKnockoff and EGKnockoff.
\begin{lemma}
	Denote the GKnockoff matrix of $\mathbf{Z}$ as $\tilde{\mathbf{Z}}$. For a symmetric matrix $\mathbf{M}$ such that $\mathbf{MZ} = \mathbf{Z}$,  $\tilde{ \mathbf{Z}}^ M = \mathbf{M}  \tilde{ \mathbf{Z}}$ is also a GKnockoff of $\mathbf{Z}$.
	\label{lemma:invariant}
\end{lemma}
\noindent
{\it Proof:} According to the construction of GKnockoff, $\tilde{\mathbf{Z}}=\mathbf{Z}\left( \mathbf{I}-\boldsymbol\Sigma^{-1} \operatorname{diag}\{\mathbf{s}\}\right)+\tilde{\mathbf{U}}\mathbf{C}$, where $\boldsymbol\Sigma = \mathbf{Z}^\top \mathbf{Z}$ and $\tilde{\mathbf{U}}$ is in the null space of $\mathbf{Z}$. Providing $\mathbf{MZ} = \mathbf{Z}$,
$$\tilde{\mathbf{Z}}^M=\mathbf{M} \tilde{ \mathbf{Z}} = \mathbf{M Z}(\mathbf{ I}-\boldsymbol\Sigma^{-1} \operatorname{diag}\{\mathbf{s}\} )+ \mathbf{M} \tilde{\mathbf{U}} \mathbf{C}=\mathbf{Z}(\mathbf{ I}-\boldsymbol\Sigma^{-1} \operatorname{diag}\{\mathbf{s}\} )+ \mathbf{M} \tilde{\mathbf{U}} \mathbf{C}.$$
In addition, by symmetry of $\mathbf{M}$, $ \mathbf{Z}^\top (\mathbf{M} \tilde{\mathbf{U}}) =\mathbf{Z}^\top \mathbf{M} ^\top\tilde{\mathbf{U}}  =(\mathbf{MZ})^\top\tilde{\mathbf{U}}=\mathbf{Z}^\top\tilde{\mathbf{U}}=0$, which implies that $\mathbf{M} \tilde{\mathbf{U}}$ is in also the null space of $\mathbf{Z}$. Hence $\tilde{\mathbf{Z}}^M$ is a GKnockoff of $\mathbf{Z}$. $\hfill\qedsymbol$

Next, we prove Theorem \ref{lemma:exchange}.  We have stated that the swapped distribution is
\begin{equation}
	[{\mathbf{X}}^{*}, \tilde{ \mathbf{X}}]_{\swap{(\mathcal{G})}}^\top {\mathbf{y}}^{*} \sim N([{\mathbf{X}}^{*}, \tilde{ \mathbf{X}}]_{\swap{(\mathcal{G})}}^\top {\mathbf{X}}^{*} \boldsymbol{\theta}_1, \sigma^2 [{\mathbf{X}}^{*}, \tilde{ \mathbf{X}}]_{\swap{(\mathcal{G})}}^\top \mathbf{M}   [{\mathbf{X}}^{*}, \tilde{ \mathbf{X}}]_{\swap{(\mathcal{G})}} ).
	\label{app:df1}
\end{equation}
Then the pairwise exchangeability would hold if only the mean and variance of swapped distribution (\ref{app:df1}) are invariant. As for the mean, the elements of $[{\mathbf{X}}^{*}, \tilde{ \mathbf{X}}]_{\swap{(\mathcal{G})}}^\top {\mathbf{X}}^{*}$ are identical to $[{\mathbf{X}}^{*}, \tilde{ \mathbf{X}}]^\top {\mathbf{X}}^{*}$ except for those computed from the $j$th column with $j \in\mathcal{G}$. 
Meanwhile, for $j\in\mathcal{G}$, we have $\theta_{1j}=0$, as $\mathcal{G} \subset  \mathcal{S}^c$. Therefore, $[{\mathbf{X}}^{*}, \tilde{ \mathbf{X}}]_{\swap{(\mathcal{G})}}^\top {\mathbf{X}}^{*} \boldsymbol{\theta}_1 = [{\mathbf{X}}^{*}, \tilde{ \mathbf{X}}]^\top {\mathbf{X}}^{*} \boldsymbol{\theta}_1$.

The variance in (\ref{app:df1}), on the other hand, can be rewritten as
\begin{equation}\label{COV}
	\sigma^2[{\mathbf{X}}^{*}, \tilde{ \mathbf{X}}]_{\swap{(\mathcal{G})}}^\top \mathbf{M}   [{\mathbf{X}}^{*}, \tilde{ \mathbf{X}}]_{\swap{(\mathcal{G})}} =\sigma^2
	\begin{pmatrix}
		&{\mathbf{X}}^{*\top}_{\swap{(\mathcal{G})}} \mathbf{M}  {\mathbf{X}}^{*}_{\swap{(\mathcal{G}})}, & {\mathbf{X}}^{*\top}_{\swap{(\mathcal{G})}} \mathbf{M}  \tilde{\mathbf{X}}_{\swap{(\mathcal{G})}} \\
		&\tilde{\mathbf{X}}^\top_{\swap{(\mathcal{G})}} \mathbf{M}  {\mathbf{X}}^{*}_{\swap{(\mathcal{G})}}, & \tilde{\mathbf{X}}^\top_{\swap{(\mathcal{G})}} \mathbf{M}  \tilde{\mathbf{X}}_{\swap{(\mathcal{G})}}
	\end{pmatrix},
\end{equation}
where $\mathbf{Z}_{\swap(\mathcal{G})}$, with slight abuse of the notation ``$\swap(\cdot)$", represents substituting the the respective columns of matrix $\mathbf{Z}$ in $\mathcal{G}$ by their counterparts. Then the $(i,j)$th entry in the first block of (\ref{COV}) is indeed
\begin{equation}\label{first-block}
	\{{\mathbf{X}}^{*\top}_{\swap{(\mathcal{G})}} \mathbf{M}  {\mathbf{X}}^{*}_{\swap{(\mathcal{G})}}\}_{(i,j)} = \begin{cases}
		X_i^{*\top} \mathbf{M}   X^*_j=X_i^{*\top}X^*_j= \mathbf{\Sigma}^*_{(i,j)}, & \text{if }i \notin \mathcal{G}, \ j \notin \mathcal{G} \\
		\tilde{X}_i^\top \mathbf{M}   X^*_j=\tilde{X}_i^\top X^*_j= \mathbf{\Sigma}^*_{(i,j)} , & \text{if }i \in \mathcal{G}, \ j \notin \mathcal{G}   \\
		X_i^{*\top}  \mathbf{M}   \tilde{X}_j =X_i^{*\top}\tilde{X}_j= \mathbf{\Sigma}^*_{(i,j)}, & \text{if }\ i \notin \mathcal{G}, \  j\in \mathcal{G}  \\	
		\tilde{X}_i^\top \mathbf{M}   \tilde{X}_j=\{{\mathbf{X}}^{*\top} {\mathbf{X}}^{*}\}_{(i,j)} = \mathbf{\Sigma}^*_{(i,j)}, &\text{if }i \in \mathcal{G}, \ j\in\mathcal{G}
	\end{cases}
\end{equation}
The equality of first three cases in (\ref{first-block}) holds attributed to that $\mathbf{M}   {\mathbf{X}}^{*} = \mathbf{M}  ^\top {\mathbf{X}}^{*} ={\mathbf{X}}^{*} $, and $\tilde{ \mathbf{X}}$ is the GKnockoff of ${\mathbf{X}}^{*}$. Note that for the second and third cases, $i\neq j$ since they belong to different sets. For the last case in (\ref{first-block}), Lemma \ref{lemma:invariant} in the appendix indicates that $\mathbf{M}   \tilde{ \mathbf{X}}$ is also a GKnockoff of ${\mathbf{X}}^{*}$, thus $	\tilde{X}_i^\top \mathbf{M}   \tilde{X}_j  = \{(\mathbf{M}   \tilde{\mathbf{X}})^\top \mathbf{M}   \tilde{\mathbf{X}}\}_{(i,j)} = \{{\mathbf{X}}^{*\top} {\mathbf{X}}^{*}\}_{(i,j)} = \mathbf{\Sigma}^*_{(i,j)}$. In sum, \begin{equation}
	{\mathbf{X}}^{*\top}_{\swap{(\mathcal{G})}} \mathbf{M}  {\mathbf{X}}^{*}_{\swap{(\mathcal{G})}} = \mathbf{\Sigma}^*={\mathbf{X}}^{*\top}{\mathbf{X}}^{*}={\mathbf{X}}^{*\top} \mathbf{M}  {\mathbf{X}}^{*}.
\end{equation}

Next, the $(i,j)$th entry in the second block of (\ref{COV}) is
\begin{equation}\label{second-block}
	\{{\mathbf{X}}^{*\top}_{\swap{(\mathcal{G})}} \mathbf{M}  \tilde{\mathbf{X}}_{\swap{(\mathcal{G})}}\}_{(i,j)} = \begin{cases}
		X_i^{*\top} \mathbf{M}   \tilde{X}_j= X_i^{*\top}  \tilde{X}_j, &\text{if }  i\notin \mathcal{G}, \ j\notin\mathcal{G} \\
		\tilde{X}_i^\top \mathbf{M}   \tilde{X}_j = \{{\mathbf{X}}^{*\top} {\mathbf{X}}^{*}\}_{(i,j)} = \mathbf{\Sigma}^*_{(i,j)}, & \text{if } i \in \mathcal{G}, \ j\notin\mathcal{G} \\
		X_i^{*\top} \mathbf{M}   X^*_j=X_i^{*\top}X^*_j= \mathbf{\Sigma}^*_{(i,j)} , &\text{if } i\notin\mathcal{G}, \ j\in \mathcal{G} \\	
		\tilde{X}_i^\top \mathbf{M}   X^*_j= \tilde{X}_i^{\top}  X_j^* , &\text{if } i\in \mathcal{G}, \ j\in\mathcal{G}
	\end{cases}
\end{equation}
The second and third cases in (\ref{second-block}) directly follow (\ref{first-block}), where $i$ and $j$ are in distinct sets hence can not be equal. In the first and last cases, by the construction rule of GKnockoffs,
\begin{equation}
	X_i^{*\top}\tilde{X}_j = \begin{cases}
		\boldsymbol\Sigma^*_{(i,j)} &\textit{if }i\neq j\\
		\boldsymbol\Sigma^*_{(i,i)}-s_i&\textit{if }i=j
	\end{cases}
\end{equation}
Therefore,
\begin{equation}
	\{{\mathbf{X}}^{*\top}_{\swap{(\mathcal{G})}} \mathbf{M}  \tilde{\mathbf{X}}_{\swap{(\mathcal{G})}}\}=\boldsymbol\Sigma^*-\diag(\mathbf{s})={\mathbf{X}}^{*\top}\mathbf{M}  \tilde{\mathbf{X}}.
\end{equation}
Similar arguments applying to the remaining two blocks of (\ref{COV}), we easily obtain
\begin{equation}
	[{\mathbf{X}}^{*}, \tilde{ \mathbf{X}}]_{\swap{(\mathcal{G})}}^\top \mathbf{M}   [{\mathbf{X}}^{*}, \tilde{ \mathbf{X}}]_{\swap{(\mathcal{G})}} = 	\begin{pmatrix}
		&\mathbf{\Sigma}^*,  &\mathbf{\Sigma}^* - \diag(\mathbf{s}) \\
		&\mathbf{\Sigma}^* - \diag(\mathbf{s}), &\mathbf{\Sigma}^*
	\end{pmatrix}=[{\mathbf{X}}^{*}, \tilde{ \mathbf{X}}]^\top \mathbf{M}   [{\mathbf{X}}^{*}\tilde{ \mathbf{X}}].
\end{equation}
The pairwise exchangeability subsequently holds.

\subsection{Proof of Theorem \ref{th1}}
In light of Theorem \ref{lemma:exchange}, we can apply Lemma 1 in \cite{Rina2015} to prove the theorem.

{
	\subsection{Proof of Theorem \ref{thm:powerG}}
	\label{app:powerG}
	
	\subsubsection{Some useful lemmas for Theorem \ref{thm:powerG}}
	
	To begin with, we define a strong signal set as $\calS_s = \{j \in \{1, \ldots,m\}: |\mathbf{d}_j^\top \bbeta | \gg  C_{\ell} s \lambda \}$, where $C_{\ell}$ is a positive constant, $s$ is the cardinality of active set $\calS = \{j \in \{1, \ldots,m\}: \bd_j^\top \bbeta \ne 0\}$, and $\lambda = C_{\lambda}  \sigma\sqrt{\log m/n}$ for a positive number $C_{\lambda}$.  All features are scaled to $\|X_j^*\|_2^2 /n = 1$ for $j  = 1, \ldots, m$. The following three conditions are imposed to ensure the power of GKnockoff and EGKnockoff. 
	
	\begin{condition}
		The matrix $ 2 \diag\{\bs \} - \diag\{ \bs\} (\bSigma^*)^{-1} \diag\{\bs\}$ is positive definite.
		\label{condition:power1}
	\end{condition}
	
	\begin{condition}
		\label{condition:power2}
		The cardinality of $\calS_s$ satisfies $|\calS_s| \ge C_{s} s$ for a constant $C_{s} \in ((2qs)^{-1}, 1)$, where $q$ is a self-specified FDR level.
	\end{condition}
	\begin{condition}\label{condition:min-signal}
		$\min_{j \in \calS} |\mathbf{d}_j^\top \bbeta | \ge 2 \kappa_n \lambda$ for some $\kappa_n \to \infty$ and $\kappa_n\lambda\to 0$ as $n \to \infty$. 
	\end{condition}
	Condition \ref{condition:power1}, \ref{condition:power2} and \ref{condition:min-signal} are common in power analyses of Knockoff frameworks in literature. See \cite{fan2019rank} and references therein. 
	Condition  \ref{condition:power2}  requires the active set to include some strong signals, and the size of strong signal set depends on the nominal FDR level $q$. A smaller value of $q$, i.e., a lower tolerance of false discovery requires stronger signals to achieve desirable power. Condition \ref{condition:min-signal} is the so-called minimum signal condition, which consists of a divergent sequence $\kappa_n$. As shown in Theorem \ref{thm:powerG}, a faster divergent speed probably leads to a higher power.
	
	We then state two useful lemmas for power analysis. Lemma \ref{lemma:consis} gives an $\ell_1$ error bound for Lasso estimate and  Lemma \ref{lemma:smallset} gives a lower bound of selection set.
	
	\begin{lemma}
		\label{lemma:consis}
		Under Condition \ref{condition:power1}, with probability at least $1 - c_{\ell_1} m^{-c_{\ell_1}}$, the $l_1$-norm error of parameters estimation $\| \hbtheta_1 - \btheta_1\|_1+ \|\tbtheta_1 \|_1 \le C_{\ell} s \lambda$.
	\end{lemma}
	
	\vspace{-0.1in}
	\noindent
	\textbf{Proof:} 	Let $\bX^*_{KO} = [\bX^*, \tbX]$, $\btheta_{KO} = [\btheta_1^\top, \mathbf{0}^\top]^\top$, and $\hbtheta_{KO} = [\hbtheta_1^\top, \tbtheta_1^\top]^\top$ be the Lasso estimate of $\btheta_{KO}$. Since $\hbtheta_{KO}$ is the minimizer of Problem \eqref{equ:aug_lasso}, it implies
	\begin{equation}
		\label{equ:consis1}
		\frac{1}{2n} \| \by^* - \bX_{KO}^* \hbtheta_{KO}  \|_2^2 + \lambda \| \hbtheta_{KO} \|_1 \le \frac{1}{2n} \| \by^* - \bX_{KO}^* \btheta_{KO} \|_2^2 + \lambda \| \btheta_{KO} \|_1.
	\end{equation}
	Note that $\by^* = \bX^*_{KO} \btheta_{KO} + \beps^*$. Pulgging it into \eqref{equ:consis1}, we get
	\begin{equation}
		\label{inequ:or1}
		\begin{aligned}
			\frac{1}{2n} \|\bX^*_{KO}(\btheta_{KO} - \hbtheta_{KO} ) + \beps^* \|_2^2 + \lambda \| \hbtheta_{KO} \|_1  \le \frac{1}{2n} \|\beps^* \|_2^2 + \lambda \| \btheta_{KO} \|_1.
		\end{aligned}
	\end{equation}
	After some calculation, we simplify inequality \eqref{inequ:or1} as
	\begin{equation*}
		\frac{1}{2} (\hbtheta_{KO} - \btheta_{KO})^\top \frac{\bX_{KO}^{* \top} \bX_{KO}^* }{n} (\hbtheta_{KO} - \btheta_{KO}) + \lambda \|\hbtheta_{KO} \|_1 \le \frac{1}{n} \beps^{* \top} \bX_{KO}^* (\hbtheta_{KO} - \btheta_{KO}) + \lambda \| \btheta_{KO} \|_1.
	\end{equation*}
	Let $\bdelta = \hbtheta_{KO} - \btheta_{KO}$ and $\mathbf{G}^* = \frac{\bX_{KO}^{* \top} \bX_{KO}^* }{n}$, the above inequality can be written as
	\begin{equation}
		\label{equ:consis2}
		\frac{1}{2} \bdelta^\top \mathbf{G}^* \bdelta + \lambda \| \hbtheta_{KO}\|_1 \le \frac{1}{n} \beps^{* \top} \bX_{KO}^*\bdelta + \lambda \|\btheta_{KO} \|_1.
	\end{equation}
	The stochastic part of inequality \eqref{equ:consis2} is bounded by
	$  \| \frac{1}{n} \bX_{KO}^{* \top} \beps^*\|_{\infty} \| \bdelta\|_1$
	according to Holder's inequality. And note that $\beps^* \sim N(\mathbf{0},\sigma^2 \bM )$, thus the covariance of $ \frac{1}{\sqrt{n}}  \bX_{KO}^{*\top} \beps^* $ equals to $\sigma^2 \bX_{KO}^{*\top} \bM \bX_{KO}^{*}/n $, and 
	$$
	\diag \{ \sigma^2 \bX_{KO}^{*\top} \bM \bX_{KO}^{*}/n \} = \diag \{ \sigma^2 \bX_{KO}^{*\top} \bX_{KO}^{*}/n \} = \sigma^2 \bI_{2m \times 2m}
	$$
	according to {Lemma \ref{lemma:invariant}} and $\|X_j^* \|_2^2 / n = 1$ for each $j$.  Therefore, each coordinate of $\frac{1}{\sqrt{n}}  \bX_{KO}^* \beps^*$ is a sub-Gaussian random variable with mean $0$ and variance $\sigma^2$. As a result, let $\lambda_0 = C_{\lambda}^\prime \sigma \sqrt{\frac{\log m}{n}}$, where $C_{\lambda}^\prime$ is a constant,
	\begin{equation*}
		\Pr \{  \| \frac{1}{n} \bX_{KO}^{* \top} \beps^*\|_{\infty} \ge \lambda_0 \} \le 4m \exp \{ - \frac{C_{\lambda}^{\prime2}}{2} \log m \}  \le    c_{\ell_1} m^{-c_{\ell_1}}
	\end{equation*}
	for some $c_{\ell_1} \ge 0$. Therefore, with probability $1 - c_{\ell_1} m^{-c_{\ell_1}}$, Inequality \eqref{equ:consis2} implies
	\begin{equation}
		\label{equ:consis3}
		\frac{1}{2} \bdelta^\top \mathbf{G}^* \bdelta  + \lambda \| \hbtheta_{KO} \|_1 \le \lambda_0 \|\bdelta\|_1 + \lambda \|\btheta_{KO} \|_1.
	\end{equation}
	Further note $ \| \hbtheta_{KO} \|_1   =  \| \hbtheta_{KO, \calS} \|_1  +  \| \hbtheta_{KO, \calS^c} \|_1 $, $\| \btheta_{KO} \|_1 = \| \btheta_{KO, \calS} \|_1$, $\| \bdelta_{\calS^c}\|_1 =  \| \hbtheta_{KO, \calS^c}\|_1$ and $\| \btheta_{KO, \calS}\|_1 - \| \hbtheta_{KO, \calS}\|_1 \le \| \btheta_{KO, \calS} - \hbtheta_{KO, \calS}\|_1 = \| \bdelta_{\calS}\|$. With some calculation, Inequality \eqref{equ:consis3} can be simplified as
	\begin{equation}
		\label{equ:consis4}
		\frac{1}{2} \bdelta^\top \mathbf{G}^* \bdelta  + \lambda \| \bdelta_{\calS^c}\|_1 \le \lambda_0 \| \bdelta\|_1 + \lambda \|\bdelta_{\calS} \|_1. 
	\end{equation}
	Let  $\lambda \ge 2 \lambda_0$, Inequality \eqref{equ:consis4} can be written as 
	\begin{equation}
		\label{equ:consis5}
		\bdelta^\top \mathbf{G}^* \bdelta  + \lambda \| \bdelta_{\calS^c}\|_1 \le 3 \lambda \|\bdelta_{\calS} \|_1.
	\end{equation}
	Therefore, we obtain $\| \bdelta_{\calS^c}\|_1 \le 3 \|\bdelta_{\calS} \|_1$. Note that both of $\bSigma^*$ and $2 \diag\{\bs \} - \diag\{ \bs\} \bSigma^{* -1} \diag\{\bs\}$ are positive definite, which yield
	$$
	\Lambda_{\min}(\mathbf{G}^*) \ge \Lambda_{\min}(\bSigma^*) \Lambda_{\min} ( 2 \diag\{\bs \} - \diag\{ \bs\} \bSigma^{* -1} \diag\{\bs\} ) \ge c_0
	$$
	for some positive $c_0$.  Therefore, according to the Inequality \eqref{equ:consis5},
	$$
	c_0 \| \bdelta \|_2^2 \le 3 \lambda \| \bdelta_{\calS}\|_1 \le 3 \lambda \sqrt{s} \| \bdelta_{\calS} \|_2 \le 3 \lambda \sqrt{s} \| \bdelta \|_2.
	$$
	Consequently,  $\|\bdelta \|_2 \le  \frac{3\lambda \sqrt{s}}{c_0}$, and
	$$
	\| \bdelta\|_1 = \|\bdelta_{\calS} \|_1 + \| \bdelta_{\calS^c}\|_1 \le 4 \|\bdelta_{\calS} \|_1  \le 4 \sqrt{s} \| \bdelta_{\calS} \|_2 \le  \frac{12 s  \lambda}{c_0} \le C_{\ell} s \lambda.
	$$
	for some $C_{\ell} \ge 12/c_0$.
	
	\begin{lemma}
		\label{lemma:smallset}
		Under Condition \ref{condition:power1} and \ref{condition:power2}, the cardinality of the selection set  $\hcalS = \{j: W_j \ge T\}$, i.e., $| \hcalS|$, is greater than $C_{s}s$ with probability $1 - c_{\ell_1} m^{-c_{\ell_1}}$.
	\end{lemma}
	\vspace{-0.1in}
	\noindent
	\textbf{Proof:} This proof is inspired by Lemma 6 of \cite{fan2019rank}. By Lemma \ref{lemma:consis},  with probability $1 - c_{\ell_1} m^{-c_{\ell_1}}$,
	$$
	|\hat{\theta}_{1j} - \theta_{1j} | \le C_{\ell} s \lambda \text{ and } |\tilde{\theta}_{1j} | \le  C_{\ell} s \lambda.
	$$ 
	Thus for each $1 \le j \le m$, we have 
	$
	W_j = | \htheta_{1j}| - | \tilde{\theta}_{1j}| \ge -| \tilde{\theta}_{1j}|  \ge - C_{\ell} s \lambda.
	$
	On the other hand, for each $j \in \calS_{s} = \{ j: | \theta_{1j}| \gg  C_{\ell} s \lambda\}$, it holds that 
	$$
	\begin{aligned}
		W_j &=  | \hat{\theta}_{1j}| - | \tilde{\theta}_{1j}| \\
		&\ge | \theta_{1j}|  - |\theta_{1j} - \hat{\theta}_{1j} | - | \hat{\theta}_{1j}| \gg C_{\ell} s \lambda.
	\end{aligned}
	$$
	which imples $W_j \gg \min_{j}W_j$ for every $j \in \calS_S$. Thus $|\{j: W_j \ge T\}| \ge |\calS_2| \ge C_{s}s$.
	
}

\subsubsection{Proof of Theorem \ref{thm:powerG}}

The proof is motivated by Theorem 3 of \cite{fan2019rank}. In this section, we assume all the features are standardized such that $\| X_j^*\|_2^2 /n = 1$ for $j = 1, \dots, m$, and we choose the coefficients difference statistics (LCD) for ease of proof. Assume there are no ties in the magnitude of nonzero $W_j's$ and no ties in the nonzero components of the Lasso solutions with asymptotic probability one. Let $|W_{(1)}| \ge \ldots \ge |W_{(m)}| $ be the ordered GKnockoff LCD statistics $W_j  = |\hat{\theta}_{1j}| -| \tilde{\theta}_{1j}|$ according to the magnitude. Denote $j^*$ as the index such that $W_{(j^*)} = T$, where $T$ is the threshold under the nominal FDR level $q $. By the definition of $T$ in \eqref{equ:T_q}, $T$ is the optimal stopping time, which means $W_{(j^*+1)} \le 0$. Therefore, it holds that $-T < W_{(j^*+1)} \le 0$. Next, we analyze two cases of $W_{(j^*+1)} =0$ and $-T < W_{(j^*+1)} <0$.

\textbf{Case 1:} Consider $-T < W_{(j^*+1)} < 0$. Since $T$ is the optimal stopping time, we have 
$$
\frac{|\{j: W_j \le -T\}| + 2}{|\{j:W_j \ge T\}|} > q,
$$
which implies $|\{j: W_j \le -T\}|  > q|\{j:W_j \ge T\}| -2$. By Lemma \ref{lemma:smallset}, $|\{j:W_j \ge T\}| \ge C_{s}s$ with high probability, thus $|\{j: W_j \le -T\}|  > qC_{s}s - 2 > 0$. Moreover, $W_j \le -T$ implies $|\htheta_{1j} | - | \ttheta_{1j} | \le -T$ and thus $ | \ttheta_{1j} | \ge T$. By Lemma \ref{lemma:consis}, with probability $1 - c_{\ell_1}m^{-c_{\ell_1}}$,
\begin{equation}
	\label{equ:power3}
	C_{\ell} s \lambda \ge \sum_{j: W_j \le -T} |\ttheta_{1j} | \ge T |\{j:W_j \le -T\}|.
\end{equation}
Combining these results, we have $C_{\ell} s \lambda \ge T(qC_{s}s - 2)$,  thus,
$$
T \le \frac{C_{\ell} s \lambda}{qC_{s}s -2} \le \kappa_n \lambda,
$$
where $C_{\ell}$, $C_{s}$ and $q$ are constants, but $\kappa_n \to + \infty$ as $n \to \infty$. 
We now control the type II error. In light of Lemma \ref{lemma:consis}, we derive
$$
C_{\ell} s \lambda \ge \sum_{j \in \calS \cap \hcalS^c } (|\htheta_{1j}  - \theta_{1j}| + |\ttheta_{1j} |) \ge \sum_{j \in \calS \cap \hcalS^c } (|\htheta_{1j}  - \theta_{1j}| + |\htheta_{1j} | - T), 
$$
since $| \htheta_{1j} | -| \ttheta_{1j}| \le T$ for $j \in \hcalS^c$. Using the triangular inequality and noting that $| \theta_{1j}| \ge 2\kappa_n  \lambda$ for $j \in \calS$,
$$
C_{\ell} s \lambda  \ge \sum_{j \in \calS \cap \hcalS^c } [| \theta_{1j}|- T] \ge ( 2\kappa_n \lambda - T) |\calS \cap \hcalS^c| \ge  \kappa_n \lambda  |\calS \cap \hcalS^c|,
$$
where the second inequality is from Condition \ref{condition:min-signal}. Thus, it follows that 
$$
\frac{|\calS \cap \hcalS|}{s} = 1 - \frac{|\calS \cap \hcalS^c|}{s} \ge 1 - \frac{C_{\ell}}{\kappa_n}.
$$

\textbf{Case 2:} In this case, $W_{(j^*+1)} = 0$ and $-T < W_{(j^*+1)}  = 0$, which means $\hcalS = \{j: W_j > 0\}$ and $\{j: W_j < -T\} = \{j: W_j < 0\}$. If $\{j: W_j < 0\} \ge q_n s$, where $q_n =  \frac{C_{\ell}}{\kappa_n}$. Then according to Inequality \eqref{equ:power3}, 
$$
C_{\ell} s \lambda \ge T | \{j : W_j < 0\}| \ge Tq_n s,
$$
which implies $T \le  \kappa_n \lambda$, then reduce to \textbf{Case 1}.

On the contrary, if $|\{j : W_j < 0\}| \le q_ns$, by $\hcalS = \supp(\mathbf{w}) \backslash \{j: W_j < 0\}$, we have 
$$
|\hcalS \cap \calS| = |\supp(\mathbf{w}) \cap \calS| - |\{j:W_j < 0\} \cap \calS| \ge |\supp(\mathbf{w}) \cap \calS|-q_n s.
$$
Let us now focus on $|\supp(\mathbf{w}) \cap \calS| $. We observe $\supp(\mathbf{w})  \supset \{1, \ldots, m\} \backslash \calS_1$ where $\calS_1 = \{j: \htheta_{1j}=0 \text{ and } \ttheta_{1j} = 0 \}$, therefore 
$$
C_{\ell} s \lambda \ge \sum_{j \in \calS \cap \calS_1} | \htheta_{1j} - \theta_{1j}| = \sum_{j \in \calS \cap \calS_1} | \theta_{1j} | \ge |\calS \cap \calS_1|  \min_{j \in \calS} | \theta_{1j}| \ge |\calS \cap \calS_1| \cdot 2 \kappa_n \lambda,
$$
which implies $ |\calS \cap \calS_1|  \le \frac{C_{\ell} s}{ 2\kappa_n }$. Further, the fact that $|\calS| = s$ entails that 
$$
|(\{1, \ldots, m\} \backslash \calS_1) \cap  \calS| \ge |\calS \cap \calS_1^c| \ge s -  |\calS \cap \calS_1|   \ge (1 - \frac{C_{\ell}}{ 2\kappa_n })s.
$$
This yields that 
$$
|\supp(\mathbf{w}) \cap \calS| \ge 	|(\{1, \ldots, m\} \backslash \calS_1) \cap  \calS|  \ge (1 - \frac{C_{\ell}}{ 2\kappa_n })s.
$$
Thus
$$
\frac{|\hcalS \cap \calS|}{s} \ge 1-  \frac{C_{\ell}}{2\kappa_n} - q_n \ge 1-    \frac{2C_{\ell}}{\kappa_n} 
$$
since $q_n =  \frac{C_{\ell}}{\kappa_n}$.
Combining the above two scenarios, we have shown that with probability one,
$$
\operatorname{Power}(q) = \E\left[ \frac{|\hcalS \cap \calS|}{|\calS|}\right] \ge  1 -    \frac{2C_{\ell}}{\kappa_n}.
$$

\subsection{Proof of Theorem \ref{th2}}
\label{app:thm2}
Denote the GKnockoff of $\begin{bmatrix}
	{\mathbf{X}}^* \\
	\mathbf{0}
\end{bmatrix}$ as  $\begin{bmatrix}
	\tilde{ \mathbf{X}}_1 \\ \tilde{ \mathbf{X}}_2
\end{bmatrix}$, then by the construction of knockoffs, we have,
\begin{equation}
	{\mathbf{X}}^{*\top}{\mathbf{X}}^* =	\begin{bmatrix}
		{\mathbf{X}}^{*\top}, \mathbf{0}^\top
	\end{bmatrix} 	\begin{bmatrix}
		{\mathbf{X}}^* \\ \mathbf{0}
	\end{bmatrix}=
	\begin{bmatrix}
		\tilde{ \mathbf{X}}_1^\top, \tilde{ \mathbf{X}}_2^\top
	\end{bmatrix} \begin{bmatrix}
		\tilde{ \mathbf{X}}_1 \\ \tilde{ \mathbf{X}}_2
	\end{bmatrix}  \triangleq \mathbf{\Sigma}^*_a
\end{equation}
and
\begin{equation}
	\begin{bmatrix}
		{\mathbf{X}}^{*\top}, \mathbf{0}^\top
	\end{bmatrix} \begin{bmatrix}
		\tilde{ \mathbf{X}}_1 \\ \tilde{ \mathbf{X}}_2
	\end{bmatrix} = {\mathbf{X}}^{*\top} \tilde{ \mathbf{X}}_1 = \mathbf{\Sigma}^*_a - \diag({\mathbf{s}})
\end{equation}

The distribution of $\begin{bmatrix}
	{\mathbf{X}}^{*\top}, & \mathbf{0}^\top \\
	\tilde{ \mathbf{X}}_1^\top, & \tilde{ \mathbf{X}}_2^\top
\end{bmatrix}\begin{bmatrix}
	\mathbf{y}^* \\ \mathbf{y}^*_a
\end{bmatrix}$ is
\begin{equation}
	\begin{bmatrix}
		{\mathbf{X}}^{*\top}, & \mathbf{0}^\top \\
		\tilde{ \mathbf{X}}_1^\top, & \tilde{ \mathbf{X}}_2^\top
	\end{bmatrix}\begin{bmatrix}
		\mathbf{y}^* \\ \mathbf{y}^*_a
	\end{bmatrix}\sim N \left( 	\begin{bmatrix}
		{\mathbf{X}}^{*\top}, & \mathbf{0}^\top \\
		\tilde{ \mathbf{X}}_1^\top, & \tilde{ \mathbf{X}}_2^\top
	\end{bmatrix} \begin{bmatrix}
		{\mathbf{X}}^{*} \\ \mathbf{0}
	\end{bmatrix}  \cdot \boldsymbol{\theta}_1, \sigma^2 \mathbf{A} \right)
\end{equation}
where
\begin{equation}
	\mathbf{A} = \begin{bmatrix}
		{\mathbf{X}}^{*\top}, & \mathbf{0}^\top \\
		\tilde{ \mathbf{X}}_1^\top, & \tilde{ \mathbf{X}}_2^\top
	\end{bmatrix} 	\begin{bmatrix}
		\mathbf{M}, &\mathbf{0} \\
		\mathbf{0}, &\mathbf{I}
	\end{bmatrix}  \begin{bmatrix}
		{\mathbf{X}}^{*}, & \tilde{ \mathbf{X}}_1 \\
		\mathbf{0}, & \tilde{ \mathbf{X}}_2
	\end{bmatrix}
\end{equation}
and the distribution of $\begin{bmatrix}
	{\mathbf{X}}^{*\top}_{\swap{(\mathcal{G})}}, & \mathbf{0}_{\swap{(\mathcal{G})}}^\top \\
	\tilde{ \mathbf{X}}_{1,\swap{(\mathcal{G})}}^\top, & \tilde{ \mathbf{X}}_{2,\swap{(\mathcal{G})}}^\top
\end{bmatrix}\begin{bmatrix}
	\mathbf{y}^* \\ \mathbf{y}_a^*
\end{bmatrix}$ is
\begin{equation}
	\begin{bmatrix}
		{\mathbf{X}}^{*\top}_{\swap{(\mathcal{G})}}, & \mathbf{0}_{\swap{(\mathcal{G})}}^\top \\
		\tilde{ \mathbf{X}}_{1,\swap{(\mathcal{G})}}^\top, & \tilde{ \mathbf{X}}_{2,\swap{(\mathcal{G})}}^\top
	\end{bmatrix}\begin{bmatrix}
		\mathbf{y}^* \\ \mathbf{y}^*_a
	\end{bmatrix}\sim N \left( \begin{bmatrix}
		{\mathbf{X}}^{*\top}_{\swap{(\mathcal{G})}}, & \mathbf{0}_{\swap{(\mathcal{G})}}^\top \\
		\tilde{ \mathbf{X}}_{1,\swap{(\mathcal{G})}}^\top, & \tilde{ \mathbf{X}}_{2,\swap{(\mathcal{G})}}^\top
	\end{bmatrix} \begin{bmatrix}
		{\mathbf{X}}^{*} \\ \mathbf{0}
	\end{bmatrix}  \cdot \boldsymbol{\theta}_1, \sigma^2 \mathbf{A}_{\swap{(\mathcal{G})}} \right)
\end{equation}
where $\mathbf{A}_{\swap{(\mathcal{G})}} $ is
\begin{equation}
	\mathbf{A}_{\swap{(\mathcal{G})}} =\begin{bmatrix}
		{\mathbf{X}}^{*\top}_{\swap{(\mathcal{G})}}, & \mathbf{0}_{\swap{(\mathcal{G})}}^\top \\
		\tilde{ \mathbf{X}}_{1,\swap{(\mathcal{G})}}^\top, & \tilde{ \mathbf{X}}_{2,\swap{(\mathcal{G})}}^\top
	\end{bmatrix} 	\begin{bmatrix}
		\mathbf{M}, &\mathbf{0} \\
		\mathbf{0}, &\mathbf{I}
	\end{bmatrix}  \begin{bmatrix}
		{\mathbf{X}}^{*}_{\swap{(\mathcal{G})}}, & \tilde{ \mathbf{X}}_{1,\swap{(\mathcal{G})}} \\
		\mathbf{0}_{\swap{(\mathcal{G})}}, & \tilde{ \mathbf{X}}_{2,\swap{(\mathcal{G})}}
	\end{bmatrix}
\end{equation}

It is trivial that the mean of distribution remains after swapping is the same if $j \in  \mathcal{S}^c$ by using the value fact of $\theta_{1,j} = 0$ for $j \in \mathcal{S}$. Next, we need to show $\mathbf{A} = \mathbf{A}_{\swap{(\mathcal{G})}}$. Define
\begin{equation}
	\mathbf{M}^\dagger = \begin{bmatrix}
		\mathbf{M}, & \mathbf{0} \\
		\mathbf{0}, &\mathbf{I}
	\end{bmatrix}.
\end{equation}
One can see that $\mathbf{M}^\dagger$ is symmetric and,
\begin{equation}
	\mathbf{M}^\dagger \begin{bmatrix}
		{\mathbf{X}}^{*} \\ \mathbf{0}
	\end{bmatrix} = \begin{bmatrix}
		\mathbf{M} {\mathbf{X}}^{*} \\
		\mathbf{0}
	\end{bmatrix}=
	\begin{bmatrix}
		{\mathbf{X}}^{*} \\ \mathbf{0}
	\end{bmatrix}
\end{equation}
then by Lemma \ref{lemma:invariant}, $\tilde{ \mathbf{X}}_a = \mathbf{M}^\dagger \begin{bmatrix}
	\tilde{ \mathbf{X}}_1 \\ \tilde{ \mathbf{X}}_2
\end{bmatrix}$ is also a GKnockoff of $ \begin{bmatrix}
	{\mathbf{X}}^{*} \\ \mathbf{0}
\end{bmatrix} $. Using this fact, we can simplified $\mathbf{A}$ as
\begin{equation}
	\label{equ:aug_G}
	\begin{aligned}
		\mathbf{A} &= \begin{bmatrix}
			{\mathbf{X}}^{*\top}, & \mathbf{0}^\top \\
			\tilde{ \mathbf{X}}_1^\top, & \tilde{ \mathbf{X}}_2^\top
		\end{bmatrix} 	\begin{bmatrix}
			\mathbf{M}, &\mathbf{0} \\
			\mathbf{0}, &\mathbf{I}
		\end{bmatrix}  \begin{bmatrix}
			{\mathbf{X}}^{*}, & \tilde{ \mathbf{X}}_1 \\
			\mathbf{0}, & \tilde{ \mathbf{X}}_2
		\end{bmatrix} \\
		&=\begin{bmatrix}
			{\mathbf{X}}^{*\top} \mathbf{M} {\mathbf{X}}^{*}, & {\mathbf{X}}^{*\top}\mathbf{M} \tilde{ \mathbf{X}}_1 \\
			\tilde{ \mathbf{X}}_1^\top \mathbf{M} {\mathbf{X}}^{*}, & \tilde{ \mathbf{X}}_1^\top \mathbf{M} \tilde{ \mathbf{X}}_1 + \tilde{ \mathbf{X}}_2^\top \tilde{ \mathbf{X}}_2
		\end{bmatrix}.
	\end{aligned}
\end{equation}
By the construction of GKnockoff, we have shown that ${\mathbf{X}}^{*\top} \mathbf{M} {\mathbf{X}}^{*} = {\mathbf{X}}^{*\top} {\mathbf{X}}^* = \mathbf{\Sigma}^*_a$,  ${\mathbf{X}}^{* \top}\mathbf{M} \tilde{ \mathbf{X}}_1 = {\mathbf{X}}^{*\top} \tilde{ \mathbf{X}}_1 = \mathbf{\Sigma} ^*_a- \diag(\mathbf{s})$ and $\tilde{ \mathbf{X}}_1^\top \mathbf{M} \tilde{ \mathbf{X}}_1 + \tilde{ \mathbf{X}}_2^\top \tilde{ \mathbf{X}}_2 = \begin{bmatrix}\tilde{ \mathbf{X}}_1^\top, \tilde{ \mathbf{X}}_2^\top\end{bmatrix} 	\begin{bmatrix}
	\mathbf{M}, &\mathbf{0} \\
	\mathbf{0}, &\mathbf{I}
\end{bmatrix}^2 \begin{bmatrix}
	\tilde{ \mathbf{X}}_1 \\
	\tilde{ \mathbf{X}}_2
\end{bmatrix} = \tilde{ \mathbf{X}}^{\top}_a \tilde{ \mathbf{X}}_a = \mathbf{\Sigma}_a^*$, therefore
\begin{equation}
	\mathbf{A}  = \begin{bmatrix}
		\mathbf{\Sigma}^*_a, &\mathbf{\Sigma}^*_a - \diag(\mathbf{s}) \\
		\mathbf{\Sigma} ^*_a- \diag(\mathbf{s}), &\mathbf{\Sigma}^*_a
	\end{bmatrix}.
\end{equation}
Then we need to show $\mathbf{A}_{\swap{(\mathcal{G})}} = \mathbf{A}$, where
\begin{equation}
	\begin{aligned}
		&	\mathbf{A}_{\swap{(\mathcal{G})}} =\begin{bmatrix}
			{\mathbf{X}}^{*\top}_{\swap{(\mathcal{G})}}, & \mathbf{0}_{\swap{(\mathcal{G})}}^\top \\
			\tilde{ \mathbf{X}}_{1,\swap{(\mathcal{G})}}^\top, & \tilde{ \mathbf{X}}_{2,\swap{(\mathcal{G})}}^\top
		\end{bmatrix} 	\begin{bmatrix}
			\mathbf{M}, &\mathbf{0} \\
			\mathbf{0}, &\mathbf{I}
		\end{bmatrix}  \begin{bmatrix}
			{\mathbf{X}}^{*}_{\swap{(\mathcal{G})}}, & \tilde{ \mathbf{X}}_{1,\swap{(\mathcal{G})}} \\
			\mathbf{0}_{\swap{(\mathcal{G})}}, & \tilde{ \mathbf{X}}_{2,\swap{(\mathcal{G})}}
		\end{bmatrix} \\
		&=\begin{bmatrix}
			&{\mathbf{X}}^{*\top}_{\swap{(\mathcal{G})}}  \mathbf{M}   {\mathbf{X}}^{*}_{\swap{(\mathcal{G})}} + \mathbf{0}_{\swap{(\mathcal{G})}}^\top \mathbf{0}_{\swap{(\mathcal{G})}}, & {\mathbf{X}}^{*\top}_{\swap{(\mathcal{G})}} \mathbf{M}   \tilde{ \mathbf{X}}_{1, \swap{(\mathcal{G})}} + \mathbf{0}_{\swap{(\mathcal{G})}}^\top \tilde{ \mathbf{X}}_{2,\swap{(\mathcal{G})}} \\
			&\tilde{ \mathbf{X}}^\top _{1, \swap{(\mathcal{G})}} \mathbf{M}   {\mathbf{X}}^{*}_{\swap{(\mathcal{G})}} + \tilde{ \mathbf{X}}_{2, \swap{(\mathcal{G})}}^\top \mathbf{0}_{\swap{(\mathcal{G})}}
			&\tilde{ \mathbf{Z}}_{1, \swap{(\mathcal{G})}}^\top  \mathbf{M}   \tilde{ \mathbf{X}}_{1, \swap{(\mathcal{G})}} + \tilde{ \mathbf{X}}_{2, \swap{(\mathcal{G})}}^\top \tilde{ \mathbf{X}}_{2, \swap{(\mathcal{G})}}
		\end{bmatrix}.
	\end{aligned}
\end{equation}
For the first block of $	\mathbf{A}_{\swap{(\mathcal{G})}}$, we see that
\begin{equation}
	{\mathbf{X}}^{*\top}_{\swap{(\mathcal{G})}}  \mathbf{M}   {\mathbf{X}}^{*}_{\swap{(\mathcal{G})}} + \mathbf{0}_{\swap{(\mathcal{G})}}^\top \mathbf{0}_{\swap{(\mathcal{G})}} =
	\begin{cases}
		{X}_i^{*\top} \mathbf{M}   {X}^*_j, &i \notin \mathcal{G}, j \notin \mathcal{G} \\
		\tilde{X}_{1,i}^\top \mathbf{M}    {X}^*_j, & i \in \mathcal{G}, j \notin \mathcal{G} \\
		{X}_i^{*\top} \mathbf{M}   \tilde{X}_{1,j},  & i \notin \mathcal{G}, j  \in \mathcal{G} \\
		\tilde{X}_{1,i}^\top\mathbf{M}   \tilde{X}_{1,j} + \tilde{ X}_{2,i}^\top \tilde{ X}_{2,j},  & i  \in \mathcal{G}, i \in \mathcal{G}
	\end{cases}
\end{equation}
Since $\mathbf{M}   {\mathbf{X}}^{*} = {\mathbf{X}}^{*}$, we have ${X}_i^{*\top} \mathbf{M}   {X}^*_j = {X}_i^{*\top} {X}^*_j =  \mathbf{\Sigma}^*_{a(i,j)}$, by the construction of $\tilde{ X}_1$ and $i \ne j$, we further conclude that $\tilde{X}_{1,i}^\top \mathbf{M}    {X}^*_j = \tilde{X}_{1,i}^\top {X}^*_j = \mathbf{\Sigma}_{a(i,j)}^* $. Recall $\mathbf{M} \begin{bmatrix}
	\tilde{ \mathbf{X}}_1 \\ \tilde{ \mathbf{X}}_2
\end{bmatrix}$ is the knockoff of $ \begin{bmatrix}
	{\mathbf{X}}^{*} \\ \mathbf{0}
\end{bmatrix}$,  therefore $	\tilde{X}_{1,i}^\top \mathbf{M}   \tilde{X}_{1,j} + \tilde{ X}_{2,i}^\top \tilde{ X}_{2,j} = \mathbf{\Sigma}^*_{a(i,j)}$. Sum it up, the first block of  $	\mathbf{A}_{\swap{(\mathcal{G})}}$ equals to $\mathbf{\Sigma}_a^*$. Using the same idea, we can show the
${\mathbf{X}}^{*\top}_{\swap{(\mathcal{G})}} \mathbf{M}   \tilde{ \mathbf{Z}}_{1, \swap{(\mathcal{G})}} + \mathbf{0}_{\swap{(\mathcal{G})}}^\top \tilde{ \mathbf{X}}_{2,\swap{(\mathcal{G})}} = \tilde{ \mathbf{X}}^\top_{1, \swap{(\mathcal{G})}} \mathbf{M}   {\mathbf{X}}^{*}_{\swap{(\mathcal{G})}} + \tilde{ \mathbf{X}}_{2, \swap{(\mathcal{G})}}^\top \mathbf{0}_{\swap{(\mathcal{G})}}  = \mathbf{\Sigma}^*_a - \diag(\mathbf{s})$, and $\tilde{ \mathbf{X}}_{1, \swap{(\mathcal{G})}}^\top  \mathbf{M}   \tilde{ \mathbf{X}}_{1, \swap{(\mathcal{G})}} + \tilde{ \mathbf{X}}_{2, \swap{(\mathcal{G})}}^\top \tilde{ \mathbf{X}}_{2, \swap{(\mathcal{G})}} = \mathbf{\Sigma}^*_a$. As a result, we have $	\mathbf{A}_{\swap{(\mathcal{G})}} = \mathbf{A}$. Then we obtain the pairwise exchangeability,
\begin{equation}
	\begin{bmatrix}
		{\mathbf{X}}^{*\top}, & \mathbf{0}^\top \\
		\tilde{ \mathbf{X}}_1^\top, & \tilde{ \mathbf{X}}_2^\top
	\end{bmatrix}\begin{bmatrix}
		\mathbf{y}^* \\ \mathbf{y}^*_a
	\end{bmatrix} \stackrel{d}{=} \begin{bmatrix}{\mathbf{X}}^{*\top}_{\swap{(\mathcal{G})}}, & \mathbf{0}_{\swap{(\mathcal{G})}}^\top \\
		\tilde{ \mathbf{X}}_{1,\swap{(\mathcal{G})}}^\top, & \tilde{ \mathbf{X}}_{2,\swap{(\mathcal{G})}}^\top
	\end{bmatrix} \begin{bmatrix}
		\mathbf{y}^* \\ \mathbf{y}^*_a
	\end{bmatrix}.
\end{equation}
By the antisymmetry and sufficiency of $\mathbf{w}$, Theorem \ref{th2} naturally follows.

{
	\subsection{Proof of  Theorem \ref{thm:powerEG}}
	\label{app:powerEG}
	It is sufficient to show $\| \hbtheta_1 - \btheta_1 \|_1 + \| \tbtheta_1\|_1 \le C_{\ell}^\prime s \lambda$ for some positive constant $C_{\ell}^\prime$. The remaining part is similar to the proof of Theorem \ref{thm:powerG}. 	Let $\bX^*_{KO} = [\bX^*_{E}, \tbX]$, $\btheta_{KO} = [\btheta_1^\top, \mathbf{0}^\top]^\top$ and $\hbtheta_{KO} = [\hbtheta_1^\top, \tbtheta_1^\top]^\top$. 	Let $\bdelta = \hbtheta_{KO} - \btheta_{KO}$ and $\mathbf{G}^* = \frac{\bX_{KO}^{* \top} \bX_{KO}^* }{n}$, obeying the same logic of Lemma \ref{lemma:consis}, we have
	\begin{equation}
		\frac{1}{2} \bdelta^\top \mathbf{G}^* \bdelta + \lambda \| \hbtheta_{KO}\|_1 \le \frac{1}{n} \beps^{* \top} \bX_{KO}^*\bdelta + \lambda \|\btheta_{KO} \|_1.
	\end{equation}
	The stochastic part $ \frac{1}{n} \beps^{* \top} \bX_{KO}^*\bdelta $ is bounded above by $ \| \frac{1}{n} \bX_{KO}^{* \top} \beps^*\|_{\infty} \| \bdelta\|_1$, where $\beps^*$ is a zero mean random vector with covariance $\sigma^2 \begin{bmatrix}
		\bM, & \mathbf{0} \\ \mathbf{0}, & \bI
	\end{bmatrix}$. Therefore, $\frac{1}{\sqrt{n}} \bX_{KO}^{* \top} \beps^*$ is a zero mean random vector with covariance
	\begin{equation*}
		\frac{\sigma^2}{n} \left[\begin{array}{cc}
			\mathbf{X}^{* \top} \mathbf{M X}^{*}, & \mathbf{X}^{* \top} \mathbf{M} \tilde{\mathbf{X}}_{1} \\
			\tilde{\mathbf{X}}_{1}^{\top} \mathbf{M X}^{*}, & \tilde{\mathbf{X}}_{1}^{\top} \mathbf{M} \tilde{\mathbf{X}}_{1}+\tilde{\mathbf{X}}_{2}^{\top} \tilde{\mathbf{X}}_{2}
		\end{array}\right]
	\end{equation*}
	according to \eqref{equ:aug_G}. Note that $\diag\{ \mathbf{X}^{* \top} \mathbf{M X}^{*}/n\} = \diag \{ \mathbf{X}^{* \top} \mathbf{X}^{*}/n\} = \bI_m$ and,
	\begin{equation*}
		\tilde{\mathbf{X}}_{1}^{\top} \mathbf{M} \tilde{\mathbf{X}}_{1}+\tilde{\mathbf{X}}_{2}^{\top} \tilde{\mathbf{X}}_{2} =  [\tbX_1^\top, \tbX_2^\top] \begin{bmatrix}
			\bM, & \mathbf{0} \\ \mathbf{0}, & \bI
		\end{bmatrix}^2 \begin{bmatrix}
			\tbX_1 \\ \tbX_2
		\end{bmatrix} = \bX^{* \top} \bX^*
	\end{equation*}
	by Lemma \ref{lemma:invariant} and $\|X_j^* \|_2^2 / n = 1$ for each $j$. As a result, $\frac{1}{n}(\diag\{  \tilde{\mathbf{X}}_{1}^{\top} \mathbf{M} \tilde{\mathbf{X}}_{1}+\tilde{\mathbf{X}}_{2}^{\top} \tilde{\mathbf{X}}_{2}  \}) = \bI_m$.   Therefore, each coordinate of $\frac{1}{\sqrt{n}}  \bX_{KO}^{*\top} \beps^*$ is a sub-Gaussian random variable with mean $0$ and variance $\sigma^2$. As a result, let $\lambda_0 = c \sigma \sqrt{\frac{\log m}{n}}$, where $c \ge 4$,
	\begin{equation}
		\Pr \{  \| \frac{1}{n} \bX_{KO}^{* \top} \beps^*\|_{\infty} \ge \lambda_0 \} \le 4m \exp \{ - \frac{c^2}{2} \log m \}  \le    c_{\ell_2} m^{-c_{\ell_2}}. 
	\end{equation}
	With this inequality, following the same routine of the  proof the {Lemma \ref{lemma:consis}}, we finally get
	\begin{equation}
		\bdelta^\top \mathbf{G}^* \bdelta  + \lambda \| \bdelta_{\calS^c}\|_1 \le 3 \lambda \|\bdelta_{\calS} \|_1.
	\end{equation}
	Together with Lemma \ref{lemma:consis} and Condition \ref{condition:power1}, we get the desired result.
}

\subsection{Proof of Theorem \ref{thm2}}
\label{sec:sis}
\subsubsection{Some useful lemmas for proving Theorem \ref{thm2}}

The following two lemmas serve as building blocks for the sure screening property of FuSIS. Lemma \ref{lemma:pearson} proves that the Pearson correlation is a U-Statistic. And Lemma \ref{lemma:inequality} studies the concentration inequality of $\hat{D}(j,h)$.

\begin{lemma}
	Denote $\{(x_i,y_i), \ i=1,\ldots,n\}$ to be a size-$n$ random sample for the random vector $(X,Y)^\top$, and $\bar{x} = \sum_{i=1}^{n}x_i/n$ and $\bar{y} = \sum_{i=1}^{n}y_i/n$ . Without loss of generality, all the samples are standardized. Then the Pearson correlation $\widehat{\Cor}(X,Y)$ is a U-Statistic.
	\label{lemma:pearson}
\end{lemma}
\noindent
{\bf Proof:} Define a kernel function
$h(x_i, x_j, y_i, y_j) =  (x_iy_i - x_iy_j + x_jy_j - x_j y_i)/2$. Then the  U-statistic specified by  function $h$ is
\begin{equation}
	U(x) = 1/C^2_n \cdot \sum_{i <j} h(x_i, x_j, y_i, y_j)
\end{equation}
Next, we show $U(x)$ is indeed the Pearson correlation.
\begin{equation}
	\begin{aligned}
		U(x) & = \frac1{n(n-1)}  \sum_{i \ne j} (x_iy_i - x_i y_j) \\
		&=  \frac1{n(n-1)}  \sum_{i,j} (x_iy_i - x_i y_j) \\
		&=  \frac1{n(n-1)}  \sum_i (nx_iy_i - x_i \sum_j y_j)\\
		& =  \frac1{n-1}  (\sum_i x_i y_i - n \bar{x} \bar{y})\\
		&=\widehat{\Cor}(X,Y)
	\end{aligned}
\end{equation}
That is, the Pearson correlation coefficient is a U-Statistic.

Lemma \ref{lemma:inequality} below provides the exponential-type deviation inequality for the screening criteria $\hat{D}(j,h)$ which is essential for the sure screening property. The corresponding population quantity is denoted by $\D(j, h) = \frac{1}{h} \sum_{i = 1}^{h} \left| \gamma_{j-i+1} - \gamma_{j+i} \right|$ where $\gamma_j$ is population covariance $\Cov(X_j, Y)$, given that $X_j$ is standardized.
\begin{lemma}
	For a given bandwidth $h$ and any $0 < \epsilon < 1$, there exists positive constants $c_1$ and $c_2$, such that
	\begin{equation}
		\operatorname{Pr} \left( | \hat{\D}(j,h) -\D(j,h) | \ge  \epsilon \right) \le hc_1 \exp \{ -c_2n \epsilon^2\}.
	\end{equation}
	\label{lemma:inequality}
\end{lemma}
\vspace{-0.4in}
\noindent
{\bf Proof:} For U-Statistics, \cite{Liu2020} has established the concentration inequality in their Lemma S.1. Together with Lemma \ref{lemma:pearson}, we easily obtain that
\begin{equation}
	\operatorname{Pr} \left( | \hat{\gamma} - \gamma|  \ge \epsilon  \right) \le c_1^\prime \exp(-c_2^\prime n \epsilon^2)
\end{equation}
where $\gamma = \Cor(X,Y)$, $c_1^\prime>0$ and $c_2^\prime>0$ are some positive constant. Note that
\begin{equation}\label{separation}
	\begin{aligned}
		\Pr \{|\hat{\D}(j,h) - \D(j,h)| \ge \epsilon \} &= \Pr\{\hat{\D}(j,h) - \D(j,h) \ge \epsilon ,\hat{\D}(j,h) \ge \D(j,h)\}\\
		&+  \Pr\{\D(j,h) - \hat{\D}(j,h) \ge \epsilon, \hat{\D}(j,h) < \D(j,h) \} \\
		& \le \Pr\{\hat{\D}(j,h) - \D(j,h) \ge \epsilon \}
		+  \Pr\{\D(j,h) - \hat{\D}(j,h) \ge \epsilon \} \\
	\end{aligned}
\end{equation}
For the first term of the right hand side in (\ref{separation}),
\begin{equation}
	\begin{aligned}
		\Pr\{\hat{\D}(j,h) - \D(j,h) \ge \epsilon \} &= \Pr\{ \frac{1}{h}\sum_{i=1}^{h}|\hat{\gamma}_{j+i-1} - \hat{\gamma}_{j-i}| - \frac{1}{h}  \sum_{i=1}^{h}|\gamma_{j+i-1} - \gamma_{j-i}|\ge \epsilon \} \\
		& = \Pr\{\sum_{i=1}^{h}(|\hat{\gamma}_{j+i-1} - \hat{\gamma}_{j-i}|-|\gamma_{j+i-1} - \gamma_{j-i}|) \ge h\epsilon \}  \\
		& \le \Pr\{\sum_{i=1}^{h}(|\hat{\gamma}_{j+i-1} - \hat{\gamma}_{j-i}-\gamma_{j+i-1} + \gamma_{j-i}|) \ge h \epsilon \} \\
		& \le \sum_{i=1}^{h}\Pr\{|\hat{\gamma}_{j+i-1} - \gamma_{j+i-1}  +  \gamma_{j-i}-\hat{\gamma}_{j-i}| \ge  \epsilon \} \\
		& \le \sum_{i=1}^{h}\Pr\{(|\hat{\gamma}_{j+i-1} - \gamma_{j+i-1} | + | \gamma_{j-i}-\hat{\gamma}_{j-i}|) \ge  \epsilon \} \\
		& \le \sum_{i=1}^{h}\left(\Pr\left\{|\hat{\gamma}_{j+i-1} - \gamma_{j+i-1} |\ge  \frac\epsilon{2}\right\} + \Pr \left\{| \gamma_{j-i}-\hat{\gamma}_{j-i}| \ge  \frac\epsilon{2} \right\}\right)  \\
		& \le 2h c_{1}^{\prime} \exp \left(-c_{2} n \epsilon^{2}\right),
	\end{aligned}
\end{equation}
where $c_2=c_2^\prime/4$. Same strategy is applied to the second term. Therefore, letting $c_1=4c_1^\prime$, we have
\begin{equation}
	\Pr \{|\hat{\D}(j,h) - \D(j,h)| \ge \epsilon \} \le hc_1 \exp(-c_2 n \epsilon^2 ).
\end{equation}

\subsubsection{Proof of Theorem \ref{thm2}}

Recall that Lemma \ref{lemma:inequality} states a exponential-type deviation inequality for the screening criteria $\hat{\D}(j,h)$
\begin{equation}
	\Pr \{|\hat{\D}(j,h) - \D(j,h)| \ge \epsilon \} \le hc_1 \exp(-c_2 n \epsilon^2 ).
\end{equation}
where $c_1$ and $c_2$ are some positive constants. We consider the complement $\mathcal{S} \nsubseteq \hat{\mathcal{A}}(\vartheta)$, meaning that there is at least one $k \in \mathcal{S}$ such that $k \in \hat{\mathcal{A}}(\vartheta)$, for the choice of $\vartheta>0$ in Theorem \ref{thm2}, we have
\begin{equation}
	\begin{aligned}
		\Pr \left( \mathcal{S} \nsubseteq \hat{\mathcal{A}}(\vartheta) \right)  = \Pr \left( \cup_{j \in \mathcal{S}} \{ j \notin \hat{\mathcal{A}}(\vartheta)  \} \right)  \le \sum_{j \in \mathcal{S}} \Pr\left( \hat{\D}(j, h) \le \vartheta \right),
	\end{aligned}
\end{equation}
Condition \ref{condition1} implies the population screening criteria for any $j \in \mathcal{S}$ has $\D(j, h) \ge 2 c_3 n^{-\kappa}$, in company with the choice of $\vartheta \le \min_{j \in \mathcal{S}} \D(j,h)/2 \le c_3 n^{-\kappa}$, we have $\D(j, h) \ge c_3  n^{-\kappa} + \vartheta$. As a result, the event $ \hat{\D}(j, h) \le \vartheta$ implies
\begin{equation}
	\D(j,h) -\hat{\D}(j, h)   \ge  c_3  n^{-\kappa}
\end{equation}
then the probability of complement event
\begin{equation}
	\begin{aligned}
		\Pr \left( \mathcal{S} \nsubseteq \hat{\mathcal{A}}(\vartheta) \right)  & \le \sum_{j \in \mathcal{S}} \Pr\left(  \D(j,h) -\hat{\D}(j, h)   \ge  c_3  n^{-\kappa}  \right) \\
		&\le \sum_{j \in \mathcal{S}} \Pr \left(| \D(j, h) - \hat{\D}(j, h)| \ge c_3  n^{-\kappa}  \right) \\
		& \le s h c_1 \exp(-c_2 c_3^2 n^{1- 2 \kappa})  = s hc_1  \exp( -c_4 n^{1 - 2 \kappa})
	\end{aligned}
\end{equation}
where $c_4 = c_2 c_3^2$, thus $\Pr \left(  \mathcal{S} \subseteq \hat{\mathcal{A}}(\vartheta) \right)  \ge 1 - O(s h \exp(-c_4n^{1 - 2 \kappa}) )$.

{
	\subsection{ Sufficient Conditions of Condition \ref{condition1}}
	\label{app:minimum_signal}
	We in this section describe some sufficient conditions to guarantee Condition \ref{condition1}, which are imposed on the covariance matrix of predictors and regression coefficients (signals). Since all features are standardized, we assume the diagonal elements of covariance matrix are all $1$ without loss of generality.
	\begin{itemize}
		\item[(C1)]\textit{(Signal strength).} Let $\bbeta_{\max} = \max_{i} |\beta_i|$, $\Delta_{\min}(\bbeta) = \min_{i,j}|\beta_{i} - \beta_{j}|I(|\beta_{i} - \beta_{j}| > 0)$. The regression coefficient satisfies $\frac{\bbeta_{\max}}{\Delta_{\min}(\bbeta)} \le c$ for a positive constant $c$ and $\Delta_{\min}(\bbeta) \ge 2c_3n^{-\kappa}$.

		\item[(C2)]\textit{(Covariance matrix).} $\bSigma_{i,j} + c \sum_{k \ne \{i,j\}} |\bSigma_{i,k} - \bSigma_{j,k} | < 1$ for any $i \ne j$, where $\bSigma$ is the population covariance matrix of the predictors $X_1,\ldots, X_p$ and $c$ is a positive constant.
		
		\item[(C3)]\textit{(Exogeneity).} $\Cov(X_j, \epsilon) = 0$ for any $j = 1,\ldots, p$, where $X_j$ is the $j$th predictor.
	\end{itemize}
	
	Condition (C1) demands the minimum difference of change positions is greater than $2c_3n^{-\kappa}$ and the maximum  value of coefficients bounded above by $c \Delta_{\min}(\bbeta)$, which could converge to zero as the sample size goes to infinity. Condition (C2), which is common in literature \citep{wang2015consistency,wang2016high}, requires the covariance matrix of predictors to be strictly diagonal dominant. Now let us prove that (C1)-(C3) sufficiently ensure Condition \ref{condition1}. 
	
	\noindent
	\textbf{Proof:} Let $\gamma_i = \Cov(X_i, Y)$ and $\bSigma_{i,j} = \Cov(X_i, X_j)$. Since $\Cov(X_i, \epsilon) = 0$, we have
	\begin{equation} 
		\begin{aligned}
			\gamma_{\tau+j} - \gamma_{\tau-{j+1}} 
			=  \Cov(X_{\tau+j}, \sum_{k=1}^p X_i \beta_i) -  \Cov(X_{\tau-j+1}, \sum_{k=1}^p X_i \beta_i)  \\
		\end{aligned}
	\end{equation}
	since $\Cov(X_i, \epsilon) = 0$. Note that $\Cov(X_{\tau+j}, \sum_{k=1}^p X_i \beta_i) = \sum_{k=1}^{p} \bSigma_{\tau+j,k} \beta_k$, and $\Cov(X_{\tau-j+1}, \sum_{k=1}^p X_i \beta_i)  = \sum_{k=1}^{p} \bSigma_{\tau-{j+1},k} \beta_k$, we obtain
	\begin{equation}
		\begin{aligned}
			\gamma_{\tau+j} - \gamma_{\tau-{j+1}} =&  \sum_{k=1}^{p} \bSigma_{\tau+j,k} \beta_k  - \sum_{k=1}^{p} \bSigma_{\tau-{j+1},k} \beta_k \\
			=&(\beta_{\tau+j} - \beta_{\tau-j+1}) - \bSigma_{\tau+j, \tau-j+1}(\beta_{\tau+j} - \beta_{\tau-j+1})  + \sum_{k \ne \{\tau+j, \tau-j+1\}} (\bSigma_{\tau+1,k} - \bSigma_{\tau-j+1,k}) \beta_k  \\
			=& (\beta_{\tau+j} - \beta_{\tau-j+1})  \Bigg(1 - \bSigma_{\tau+j, \tau-j+1} + \sum_{k \ne \{\tau+j, \tau-j+1\}} (\bSigma_{\tau+1,k} - \bSigma_{\tau-j+1,k}) \frac{\beta_k}{\beta_{\tau+j} - \beta_{\tau-j+1}} \Bigg). \\
			:=&(\beta_{\tau+j} - \beta_{\tau-j+1}) \cdot A.
		\end{aligned}
	\end{equation}
	The second equality holds because the diagonal elements of covariance matrix are 1. Note that (C1) and (C2) imply
	\begin{equation}
		A \ge 1 - \bSigma_{\tau+j, \tau-j+1} - c\sum_{k \ne \{\tau+j, \tau-j+1\}}  |\bSigma_{\tau+1,k} - \bSigma_{\tau-j+1,k}| >0 
	\end{equation}
	Obviously $A \le 1$, Therefore,  $|\gamma_{\tau+j} - \gamma_{\tau-{j+1}} |\ge |\beta_{\tau+j} - \beta_{\tau-j+1}| \ge 2c_3n^{-\kappa}$, thus $ \D(\tau,h) =\frac{1}{h} \sum_{i=1}^h |\gamma_{\tau-i+1} - \gamma_{\tau+i}| \ge 2c_3n^{-\kappa}$.
	
}

\subsection{Proof of Theorem \ref{thm:fdr}}
Theorem \ref{thm:fdr} trivially holds on the basis of Theorem \ref{th1} and \ref{thm2}.

\bibliography{reference}
\bibliographystyle{chicago}
\end{document}